\documentclass[fleqn,usenatbib,usedcolumn]{mnras}
\pdfoutput=1
   \usepackage[british]{babel}             
   \usepackage{newtxtext}                  
   \usepackage[slantedGreek]{newtxmath}    

   \usepackage[T1]{fontenc}                

\usepackage{amsmath}
\usepackage{amssymb}
\usepackage{amsfonts}
\usepackage{amstext}

\allowdisplaybreaks

\usepackage{ifthen}
\usepackage{graphicx}
\usepackage{dcolumn}
\usepackage{bm}
\usepackage{color}

\usepackage{bbold}
\usepackage{lscape}

\usepackage{fancybox}
\usepackage{pstricks}
\usepackage{soul, color}
\usepackage{latexsym}
\usepackage{pifont}
\usepackage{shorttoc}
\usepackage{subfigure}
\usepackage{textcomp} 
\usepackage{stfloats}
\usepackage{lipsum}
\usepackage{lscape}

\newcommand{\kms}{\, {\rm km\, s}^{-1}}

\newcommand{\lya}{Ly$\alpha$\ }

\newcommand{\cm}{{\rm cm}}

\def\aap{A\&A} 
\def\araa{ARAA} 
\def\aj{AJ} 
\def\mnras{MNRAS}
\def\apj{ApJ}
\def\apjs{ApJS}
\def\apjl{ApJL}

\def\aj{AJ}
\def\araa{ARAA}
\def\jcap{JCAP}

\newcommand{\angs}{\, {\textup{\rm\AA} }}
\newcommand{\no}[1]{}

\begin{document}

\title{A metal-line strength indicator for Damped Lyman Alpha (DLA) systems
 at low signal-to-noise}

\author[A. Arinyo-i-Prats et al.]{
Andreu Arinyo-i-Prats,$^{1}$\thanks{E-mail: andreuaprats@icc.ub.edu}
	Llu\'\i s Mas-Ribas,$^{2,1}$\thanks{E-mail: l.m.ribas@astro.uio.no}
	Jordi Miralda-Escud\'e,$^{3,1}$\thanks{E-mail: miralda@icc.ub.edu}
	\newauthor
	Ignasi P\'erez-R\`afols$^{4}$\thanks{E-mail:iprafols@icc.ub.edu}
	and Pasquier Noterdaeme$^{5}$\thanks{E-mail: noterdaeme@iap.fr}
\\
$^{1}$nstitut de Ci\`encies del Cosmos, Universitat de Barcelona/IEEC,
Barcelona 08028, Catalonia\\
$^{2}$Institute of Theoretical Astrophysics, University of
Oslo,Postboks 1029, 0315 Oslo, Norway\\
$^{3}$Instituci\'o Catalana de Recerca i Estudis Avan\c{c}ats,
Barcelona, Catalonia\\
$^{4}$Laboratoire d'Astrophysique de Marseille, Marseille, France\\
$^{5}$Institut d'Astrophysique de Paris, UPMC $\&$ CNRS, F-75014   Paris, France
}

\maketitle


\begin{abstract}
\label{sec:Abs}
 The Baryon Oscillation Spectroscopic Survey of SDSS-III has provided an
unprecedentedly large sample of Damped \lya systems (DLAs), the largest
repositories of neutral hydrogen in the Universe. This DLA sample has been used
to determine the DLA bias factor from their cross-correlation with the \lya
forest absorption in \cite{FontRibera2012,Perez2018}, showing that DLAs are
associated with relatively massive halos. However, the low resolution and
signal-to-noise of BOSS spectra do not allow precise measurements of the DLA
metal lines. We define a metal strength parameter, $S$, based on combining
equivalent widths of 17 metal lines, that can be measured with an optimal
signal-to-noise ratio for individual DLAs in BOSS spectra, allowing for the
classification of the DLA population into subgroups of different $S$. We present
the distribution of this DLA metal strength and the dependence of its mean value
on $N_{\rm HI}$ and redshift. We search for systematic effects and variations in
the catalogue purity by examining the dependence of the $S$ distribution on the
spectral signal-to-noise and the estimated error on $S$. A catalogue of DLAs
with measured equivalent widths for the selected 17 metal lines and the value of
$S$ is made publicly available, which will be used to measure the dependence of
the DLA bias factor on the $S$ parameter. The relation of the metal strength on
the gas metal abundances and velocity dispersion can be constrained by studying
the stacked metal absorption spectra of DLAs as a function of $S$, allowing for
future determinations of the dependence of the bias factor on the metallicity
and velocity dispersion of DLAs.

\end{abstract}

\section{Introduction}
\label{sec:intro}

  ~\par Observations of Damped Lyman Alpha systems (DLAs) in quasar absorption
spectra are crucial to understand galaxy formation at different epochs
(e.g., \citet{Wolfe1986, Wolfe+2005}).
Most of the atomic hydrogen in the Universe is present in DLAs, or
absorption systems with $N_{\rm HI} > 2\times 10^{20} \cm^{-2}$, which
are self-shielded against the external ionizing background. This atomic
hydrogen accounts for $\sim 2$ to 3\% of all the baryonic matter in the
Universe, a fraction that remains roughly constant during the epoch of
the maximum rate of star formation in the Universe at $2<z<4$
\cite{Noterdaeme2012} and is comparable to the fraction of baryons in
stars at that epoch \cite{Shapley2011}.

  A moderately high large-scale bias factor for DLAs of $b_{\rm DLA}
\simeq 2$ has been measured by \cite{FontRibera2012,Perez2018} at
$z\simeq 2.3$, implying that most DLAs are hosted by halos of massive
galaxies or galaxy groups, and that an extended distribution of atomic
gas is generally present in these halos out to radii larger than the
size of visible galaxies to account for the large cross sections
required to explain the observed rate of incidence. Models of galaxy
formation require strong winds to explain this high bias factor, which
can expel the gas from low mass halos to reduce the contribution of
dwarf galaxies to DLAs, and spread the gas out to large radius in
massive halos to increase their DLA cross section
(e.g., \citet{Barnes2014,Bird+2015}).

  Although the \lya absorption profile of a DLA tells us only the
hydrogen column density, much more information can be obtained from the
DLA metal lines: the abundances are typically similar to those of
metal-poor globular clusters and halo stars in the Milky Way, and the
absorption line profiles reveal a rich and diverse velocity structure
characterized by multiple components, with a total velocity width
ranging from $10 \kms$ to $200 \kms$
\citep{Prochaska+1997,Prochaska+1998,Wolfe1998}. The derived
metallicities are distributed over a broad range ($10^{-3}\,\leq\,
Z/Z_{\odot}\,\leq\,10^{-1}$), and their average declines slowly with
redshift (e.g., \citet{Prochaska+2002,Rafelski+2012,Rafelski2014}). 

  Most studies of the DLA metal lines are based on high-resolution, high
signal-to-noise quasar spectra, which are necessary to detect the small
equivalent width lines and to reveal the small-scale complexity of the
velocity profiles. However, the large samples of DLAs required to
measure cross-correlations that are used to infer the DLA linear bias
factor can only be obtained at present with low resolution and low
signal-to-noise spectra similar to those of the Baryon Oscillation
Spectroscopic Survey (BOSS) of SDSS-III
\cite{Eisenstein2011,Dawson2013,Smee2013,Alam+2015}.

  Our aim in this work is to define a parameter characterizing the
strength of the metal lines of a DLA, which can be measured with the
highest possible signal-to-noise ratio by combining all the metal
absorption lines that are usually observable for DLAs. If this
parameter can be measured sufficiently accurately even in low
signal-to-noise spectra similar to BOSS, then the DLAs can be classified
into groups of different metal strength, for which several average
properties such as the large-scale bias factor can be measured as a
function of the metal strength. We may hope in this way to test, for
example, the presence of a relation between host halo mass and
gas metallicity in a DLA, which is naturally expected because of the
well-known mass-metallicity relation for galaxies. As we shall explain
below, caution is required to reach any such interpretations because
the metal strength parameter we will define depends only on the metal
line equivalent widths that are measurable in low-resolution absorption
spectra, which are affected both by metal abundances and the gas
velocity dispersion. Nevertheless, the mean absorption spectrum can
also be measured for groups of DLAs with different metal strength using
the technique discussed in \cite{lluis2014}, which contains information
that can help disentangle variations of the gas velocity dispersion and
metal abundances with metal strength.

  This work is organized as follows: \S \ref{sec:data} specifies the DLA
catalogue and the quasar spectra from BOSS that we use. In \S
\ref{sec:methodology} we describe in detail our method to measure line
equivalent widths for a set of 17 low-ionization metal lines that have
the highest equivalent widths in DLAs. In \S \ref{sec:sel_par} we
define the metal strength, and a second quantity that is corrected for
the effect of the hydrogen column density on the mean strength of metal
lines. Results are presented in \S \ref{sec:results}, where we
discuss the distribution of this metal strength parameter and several
systematic effects due to impurities in the DLA catalogue, and we make
publicly available a catalogue with our measurements of equivalent widths
and the metal strength parameter for each DLA. Finally,
we present the conclusions in \S \ref{sec:conclude}. Applications
to measure the mean bias factor and other properties as a function of
metal strength will be presented in future papers.

\section{Data sample }
\label{sec:data}

  ~\par We use the quasar spectra in the complete SDSS-III BOSS Data Release
12 (DR12), from the Quasar Catalogue DR12Q \cite{Paris2016}. A detailed
description of the SDSS telescope and the BOSS instrument obtaining the
spectra is found in \cite{Gunn1998, Gunn2006, Dawson2013, Smee2013}, and
the method to select quasar targets is described in \cite{Ross2012}.
 
  We use the DR12 extension of the DLA catalogue of \cite{Noterdaeme2012},
containing a total of $34,050$ DLA candidates with a column density
$\log(N_{\rm{HI}}) \ge 20 $. The method to detect these DLAs, described in
\cite{Noterdaeme2009}, is an automatic profile recognition
procedure using Spearman correlation analysis with a Voigt profile.
Only the \lya absorption line is considered to decide if an absorption
feature is included as a DLA in the catalogue. This ensures that there is
no selection bias in favor of DLAs with strong metal lines. The presence
of metal lines is, however, used to refine the accuracy of the measured
redshift of the detected DLAs. This catalogue will be referred to 
as DR12-DLA from now on.

  The DLA sample we use in this work is very similar to the
sample of \cite{lluis2014} (who used an earlier version of the
DR12-DLA catalogue with minor differences in the selected DLAs), who
showed the distribution in redshift and
$N_{\rm HI}$ in their figure 1. Most DLAs are at redshifts $1.9<z<3.5$,
with a small fraction extending to higher redshifts. Although DLAs are
usually defined as having $\log(N_{\rm HI}) > 20.3$ (roughly the column
density at which the gas becomes mostly neutral due to self-shielding,
although this depends on the gas density; see \citet{Wolfe1986}),
we use all the systems going down to column densities
$N_{\rm HI} > 10^{20} \cm^{-2}$ to have a larger sample.

  We note that these systems are only DLA candidates, and
that some fraction of them are expected to be false DLAs arising from a
combination of noise and regions of strong \lya forest absorption that
are confused with a DLA in low signal-to-noise spectra. The purity of
the catalogue, or fraction of DLAs that are real, is expected to decline
as the signal-to-noise ratio and the column density decrease.

  The catalogue gives, for each detected DLA, the quasar and DLA
redshift, the continuum-to-noise ratio (hereafter, CNR) of the spectrum
in the \lya forest region (as defined in \cite{Noterdaeme2009}), and the
DLA column density. For each DLA in the catalogue, we use the
corresponding quasar spectrum from BOSS to measure the associated metal
lines. We use the co-added spectra, which are provided in wavelength
bins of width $\Delta \log_{10}(\lambda)=10^{-4}$, corresponding to
a velocity width of $69.05 \kms$. 

\no{ Spectral contamination by skylines is masked using the list of lines
as provided in \cite{SkyLin2014}. Also, we mask the strongest emission
features of the red part of the quasar continuum to avoid using them
when normalizing the spectrum, any DLA absorption  line close to these quasar emission lines
will not be measured. The emission features, shown in 
Table \ref{tab:em_lin}, are the quasar emission lines provided in \cite{Paris2012} Table 3.
From Table 2 of \cite{Prochaska2001} we use the laboratory wavelength 
of the 17 metal lines used in this work.}

\section{Measuring metal line  equivalent widths}
\label{sec:methodology}

  ~\par In this section we describe the method used to compute equivalent
widths ($W$) of a selection of 17 metal lines of DLAs. A weighted average of
these equivalent widths will be used as a definition of the {\it metal
strength} of each DLA. Briefly, our method uses a measurement window
around the central wavelength of the absorption line over which the
equivalent width is integrated, and two windows around it to determine a
continuum from a linear regression of the measured flux. In \S
\ref{subs:metal_lin} we explain the selection of the 17 lines. The
windows are described in \S \ref{subs:windows}. In \S
\ref{subs:select_met} we specify which of the 17 lines are used for each
individual DLA depending on how the windows are placed in the quasar
spectrum, and \S \ref{subs:continuum} gives the details of how the
equivalent widths and their errors are calculated.
 
 The DR12-DLA catalogue of \cite{Noterdaeme2009} includes measurements
of $W$ for 10 metal lines. However, their method provides a biased
estimate of these equivalent widths because they are only measured when
their detection is considered significant. This means that a metal line
may be included in the catalogue when the value of $W$ has been increased
by noise, or may be dropped when it has been reduced, systematically
affecting the average $W$. Negative equivalent widths caused by noise
also need to be included in the catalogue to avoid bias. Our method will
select the lines to be measured depending only on the location of the
windows to be used for the continuum determination and the equivalent width
integration, but not on the value of $W$ that is derived.

\subsection{List of Metal lines included}
\label{subs:metal_lin}

 The 17 metal lines we use to evaluate our metal strength for each DLA
are selected to be transitions of low-ionization species that lie in the
DLA rest-frame wavelength interval between $1260\angs$ and
$3000\angs$, with a mean equivalent width measured from the
stacked spectrum in \cite{lluis2014} of $\overline W > 0.05\angs$. The
wavelength lower limit is set to avoid the \lya forest and the \lya quasar
emission line, and the upper limit is determined by the maximum
wavelength reached by the BOSS spectrograph \cite{Smee2013} at the
lowest redshifts at which DLAs are found, combined with the absence of
strong absorption lines beyond our longest wavelength line, MgI at
$2853\angs$.

\begin{table*}
 \centering
 \caption{\label{tab_lines} Metal lines and their principal
characteristics used for this work. {\it First column:} Name of the
metal line. {\it Second column:} Wavelength in Angstroms
\citep[from Table 2 of][]{Prochaska2001}. The first
three lines are blends of two lines and we give the wavelength of the
strongest line, except for O\thinspace I-Si\thinspace II (which are comparably strong) where
we give the average of the two wavelengths. In the third blend, C\thinspace II$*$
is a metastable state of \thinspace CII. {\it Third column:} Equivalent width
measured from the stacked spectrum in \citep{lluis2014}.
{\it Fourth column:} Mean equivalent width $\overline W$ calculated with our
method as described in \S \ref{sec:methodology} for spectra with CNR $>$ 2
 as in \citep{lluis2014}. {\it Fifth column:}
 Mean contribution $C_k$ of each line to
the total metal strength in the DLAs where $W$ of the line is measured
(see \S \ref{sec:sel_par} for details). {\it Sixth column:} Fraction
$x_k$ of the $34,050$ DLAs in which each line is actually
measured. {\it Seventh column:} Number of
DLAs in which each line has been measured.}
\vspace{4 mm}
	\begin{tabular}{cc|cc|ccc}

	 \hline	
  	
	\rule{0pt}{3ex} Name & $\lambda \, [\angs]$  & Stacked $W$ $[\angs ]$ & $\overline{W}$ $[\angs ]$ & $C_k$ & $x_k $ & $N_k$  \\
	\hline
	\hline
	Si\thinspace II-Fe\thinspace II$\,$1260   &	1260.42	&	  $0.623 \pm 0.008$ &  $ 0.693 \pm 0.010	$  & 0.58 & 0.18 & 6057	\\
	O\thinspace I-Si\thinspace II$\,$1303 	  &	1303.20	&	  $0.793 \pm 0.012$ &  $0.818 \pm 0.012$   & 0.46 & 0.23 & 7807	\\
	C\thinspace II-C\thinspace II$*\,$1334    &	1334.53	&	  $0.630 \pm 0.007$ &  $ 0.596 \pm 0.006$ & 0.40 & 0.47 & 16020	\\
	Si\thinspace II$\,$1526  	  &	1526.71	&	   $ 0.443 \pm 0.004$ &  $0.396 \pm 0.005$    & 0.28 & 0.71 & 24140 	\\
	Fe\thinspace II$\,$1608 	  &	1608.45	&	   $0.228 \pm 0.004$  &  $0.218 \pm 0.004$    & 0.11 & 0.71 & 24013\\
	Al\thinspace II$\,$1670  	  &	1670.79	&	    $0.452  \pm 0.005$ &  $0.41 \pm 0.005$     & 0.27 & 0.70 & 23940	\\
	Si\thinspace II$\,$1808  	  &	1808.01	&	    $0.059 \pm  0.008$ &  $0.0575 \pm 0.005$  & 0.015 & 0.63 & 21608\\
	Al\thinspace III$\,$1854 	  &	1854.72	&	    $0.117  \pm 0.006$ &  $0.118 \pm 0.006$  & 0.031 & 0.60 & 20317 \\
	Al\thinspace III$\,$1862 	  &	1862.79	&	    $0.067 \pm  0.006$ &  $0.095 \pm 0.006$  & 0.020 & 0.59 & 20131\\
	Fe\thinspace II$\,$2344    &		2344.21	&     $0.520 \pm  0.014$	&  $0.404 \pm 0.014$   & 0.103 & 0.17 & 5732 \\
	Fe\thinspace II$\,$2374    &		2374.46	&     $ 0.282 \pm  0.014$ &  $0.278 \pm 0.018$  & 0.034 & 0.15 & 5024 \\
	Fe\thinspace II$\,$2382    &		2382.76	&     $0.67 \pm 0.03 $   &     $0.623 \pm 0.019$   & 0.15 & 0.14 & 4907 \\
	Fe\thinspace II$\,$2586    &		2586.65	&	   $ 0.46 \pm  0.02 $ &     $0.47 \pm 0.03$     & 0.06 & 0.11 & 3597 \\
	Fe\thinspace II$\,$2600    &		2600.17	&	   $0.72  \pm 0.02$  &      $0.61 \pm 0.03$     & 0.10 & 0.10 & 3458 \\
	Mg\thinspace II$\,$2796 	  &	2796.35	&	   $1.15 \pm  0.03$  &      $1.13 \pm 0.05$     & 0.18 & 0.047 & 1598 \\
	Mg\thinspace II$\,$2803 	  &	2803.53	&	   $1.07 \pm  0.03$  &      $1.02 \pm 0.05$     & 0.15 & 0.048 & 1629  \\
	Mg\thinspace I$\,$2852  	  &	2852.96	&	   $0.23 \pm  0.03$ &       $0.31 \pm 0.04$     & 0.021 & 0.040 & 1370  \\
	
	\hline
	\end{tabular}
\end{table*}

 The low-ionization species are either neutral atoms or ions that are
once or twice ionized, but we exclude higher ionization lines like
C\thinspace IV and Si\thinspace IV.
We use this combination of low-ionization transitions to obtain a metal
strength that reflects a property of the low-ionization gas with the
highest possible signal-to-noise, by combining $W$ measurements of all
the available lines. We exclude high-ionization lines because these are
known to reflect a more extended gas distribution with different
physical properties. In practice, we know that this metal strength will
depend both on the metal abundances and velocity dispersion, because of
line saturation effects.

 The 17 selected line transitions are listed in Table \ref{tab_lines},
where we give the equivalent width computed from the stacked DR12
spectrum designated as {\it total} in \cite{lluis2014}, compared to the
mean equivalent width $\overline W$ measured with our method, as described
below. The first three lines in Table \ref{tab_lines} are blends of
lines that are too close to be measured separately, and we list the
total equivalent widths for the blend. Their wavelengths are that of the
Si\thinspace II transition in the first blend (which is much stronger than Fe\thinspace II),
the average of the O\thinspace I and Si\thinspace II transitions in the second blend (which
are comparably strong), and the C\thinspace II transition in the third blend (which
is much stronger than the transition of the metastable state C\thinspace II$*$).
The rest of the quantities will be discussed in \S \ref{sec:sel_par}.


\subsection{Measurement and Continuum Windows}
\label{subs:windows}

  We use a fixed measurement window to integrate the equivalent width
for all the lines, with a width that is normally set to 15 pixels in the
co-added spectra, except for the O\thinspace I-Si\thinspace II blend for
which we use a window width of 27 pixels. The measurement window is
centered on the pixel that includes the central wavelength of the line
listed in Table \ref{tab_lines} multiplied by $1+z_d$, where $z_d$ is
the DLA redshift, and includes 7 pixels on each side (13 for O\thinspace
I-Si\thinspace II). The width of 15 pixels, corresponding to a velocity
width $1036\kms$, is generally wide enough to include most of the
absorption components in DLAs, as can be seen from the metal line
profiles in the stacked absorption spectrum of DLAs in \cite{lluis2014}
(see their Figure 10), which have a dispersion of $\sim 100$ to
$150\kms$ (only moderately wider than the Point Spread Function of the
BOSS spectrograph wavelength resolution). However, the equivalent widths
we measure are underestimated if there are large DLA redshift errors
in the DLA catalogue we use, which shift the metal lines partly outside
the measurement window.
In addition, the O\thinspace I-Si\thinspace II blended line requires the
wider window mentioned above because the two lines are resolved and are
spread over a wider interval than the other blends. We will show below
the impact on the mean equivalent widths of decreasing the measurement
window width to 10 pixels or increasing it to 20 or 25 pixels 
(Figure \ref{fig:mean_W_hist_win} and \S \ref{sec:results}). For these
cases, we also change the O\thinspace I-Si\thinspace II blend to 25, 29
and 31 pixels, respectively.

  The windows for determining the continuum are both set to a width of
21 pixels, on the left and the right of the measurement window.
We generally leave 1 pixel that is not used between the end of the
measurement window and the start of each continuum window, with the
exception of some lines for which the continuum window placed in this
way would include another DLA metal line that would systematically lower
the estimated continuum. These exceptions are listed in Table
\ref{tab:range}, where we give the space left between the center of the
line and the start of the two continuum windows on each side for these
set of lines (for all other lines, the normal space left is 8.5 pixels,
corresponding to half of the 15 pixel width of the measurement window
plus one). The size of the two continuum windows remains fixed at 21
pixels for all these lines.
  
  \begin{table}
\centering
\caption{\label{tab:range}
 Range around each metal absorption line at which the two continuum
windows start, given in pixel numbers ($p$) and velocity interval ($s$).
The two continuum windows of any of the two lines in each of the 4 pairs
start a number $p$ of pixels to the right and to the left of the line
center. When measuring one line in any of these four pairs, the other
line is avoided by leaving it in between the integrating and the
continuum windows. }
 \vspace{3 mm}
	
	\begin{tabular}{cccc}
	\hline
	
	\rule{0pt}{3ex} Name  & $\lambda [\angs]$  & $s$ [km/s] & $p$   \\
	\hline
	\hline
	Al$\,$III  - Al$\,$III &   1854.72 - 1862.79   & 2071.5   & 30     \\
	Fe$\,$II - Fe$\,$I  &    2374.46 - 2382.76  & 1726.3 & 25  \\ 
	Fe$\,$II - Fe$\,$II   &    2586.65 - 2600.17  & 2071.5  & 30  \\ 
	Mg$\,$II - Mg$\,$II  &   2796.35 - 2803.53  & 1553.6  & 22.5  \\ 
	\hline
	\end{tabular}
\end{table}
  

\subsection{Excluded metal lines for each DLA}
\label{subs:select_met}

  DLA metal lines often fall in regions of the spectrum where the
equivalent width measurement is subject to large systematic errors, due
to the presence of quasar emission lines or of sky
lines from the atmosphere that increase the noise. To reduce these
systematics we measure only a subset of the 17 metal lines for each
DLA that satisfy the following criteria:

\begin{itemize}
\item The right end of the continuum window (at the longest wavelength)
must be below the longest wavelength of the BOSS spectrum.

\item The left end of the continuum window must lie at least 30 pixels
redwards of the center
of the \lya emission line of the quasar (at $\lambda=1215.6\angs$), to
avoid the \lya forest and the effect of the \lya quasar emission line on
the measurement of $W$.

\item The left and right ends of the continuum windows must be separated
by more than 30 pixels (or $2071.5 \kms$) from the center of any of the
following four quasar emission lines: Si\thinspace IV-O\thinspace IV at
$\lambda= 1400.0\angs$, C\thinspace IV at $\lambda=1549.2\angs$,
C\thinspace III at $\lambda=1908.7\angs$, and Mg\thinspace II at
$\lambda=2798.7\angs$. These are the strongest quasar emission lines
(see, e.g., Table 3 in \cite{Paris2012}), which we avoid because they
affect the continuum determination and the measurement of $W$.

\item The measurement window must not include any skyline. We discard
the metal line whenever the central wavelength of any skyline is within
the measurement window or in one of the two adjacent pixels, to avoid
the increased systematic error on the equivalent width caused by
skylines. Following previous analyses of BOSS quasar spectra, we use the
set of 872 skylines\footnote{\url{ https://github.com/igmhub/picca/blob/master/etc/dr14-line-sky-mask.txt }}.
\end{itemize}


\subsection{Calculation of the metal line equivalent width and error} 
\label{subs:continuum}

  The quasar continuum around a metal absorption is estimated by doing a
linear regression of the flux values $f_i$ in every pixel $i$ in the two
continuum windows to the left and right of the measurement window, which
contain a total of 42 pixels.
First a linear regression\footnote{\url{ http://www.statsmodels.org/stable/generated/statsmodels.regression.linear_model.WLS.html}} 
is computed applying the inverse variance
$w_i=1/\sigma_i^2$ of the pixels $i$ as weights, where $\sigma_i$ is the
noise.
This yields a preliminary determination of a continuum $c_{i}$, from
which we compute the transmission fraction $F_{pi}=f_i/C_{pi}$.
Next, we eliminate outliers in the continuum windows, which may result
from highly noisy flux measurements in skylines. Note that DLA metal
lines are discarded only when skylines are present in the measurement
window, but not in the continuum window. We eliminate any pixels for
which $| F_{pi} - 1 | > 4$ as outliers. We repeat the determination of
the continuum and we check again for outliers, until none are left. If
this process leaves less than 5 pixels in any of the two continuum
windows, the line is discarded and not measured for the DLA being
analysed.

  Then we repeat the determination of the continuum with
\emph{non-weighted} linear regression with the pixels that are left
after elimination of outliers. The reason for not using weights in our
final determination of the continuum is that the weights that minimize
the impact of more noisy pixels also introduce a bias that
systematically lowers the continuum, because pixels with less flux are
assigned lower noise owing to the photon noise contribution calculated
by the BOSS pipeline. This
systematic effect causes equivalent widths to be underestimated due to
the lowering of the continuum, by an amount that increases with spectral
noise and introduces an unwanted artificial correlation of our metal
strength parameter and its error, so we decided not to use weights to
obtain the final continuum. With the final continuum estimation $C_i$,
we compute the transmission fraction  $F_i=f_i/C_i$.
At the same time, not using the weights for our final continuum implies
that outliers caused by skylines or other large errors can strongly 
distort the determination of the continuum and they need to be
eliminated. 

 We checked that different ways of computing the linear regression, weighted
or unweighted, and varying the outlier condition to
$| F_{pi} - 1 | > (2, 3, 5)$, have only percent effects on the mean
equivalent widths.

\begin{figure}

\centering

\includegraphics[width=0.95\columnwidth]{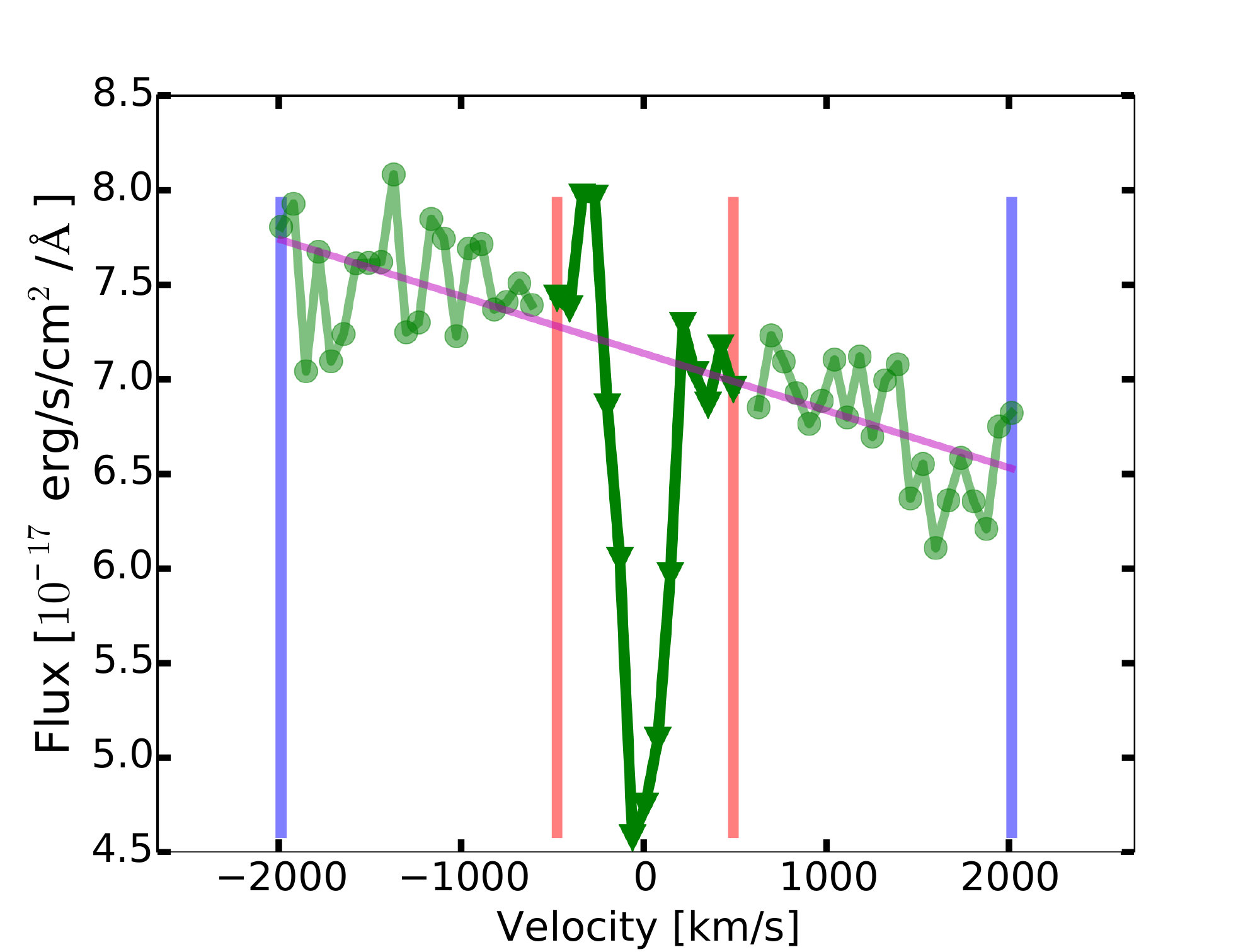}
\includegraphics[width=0.95\columnwidth]{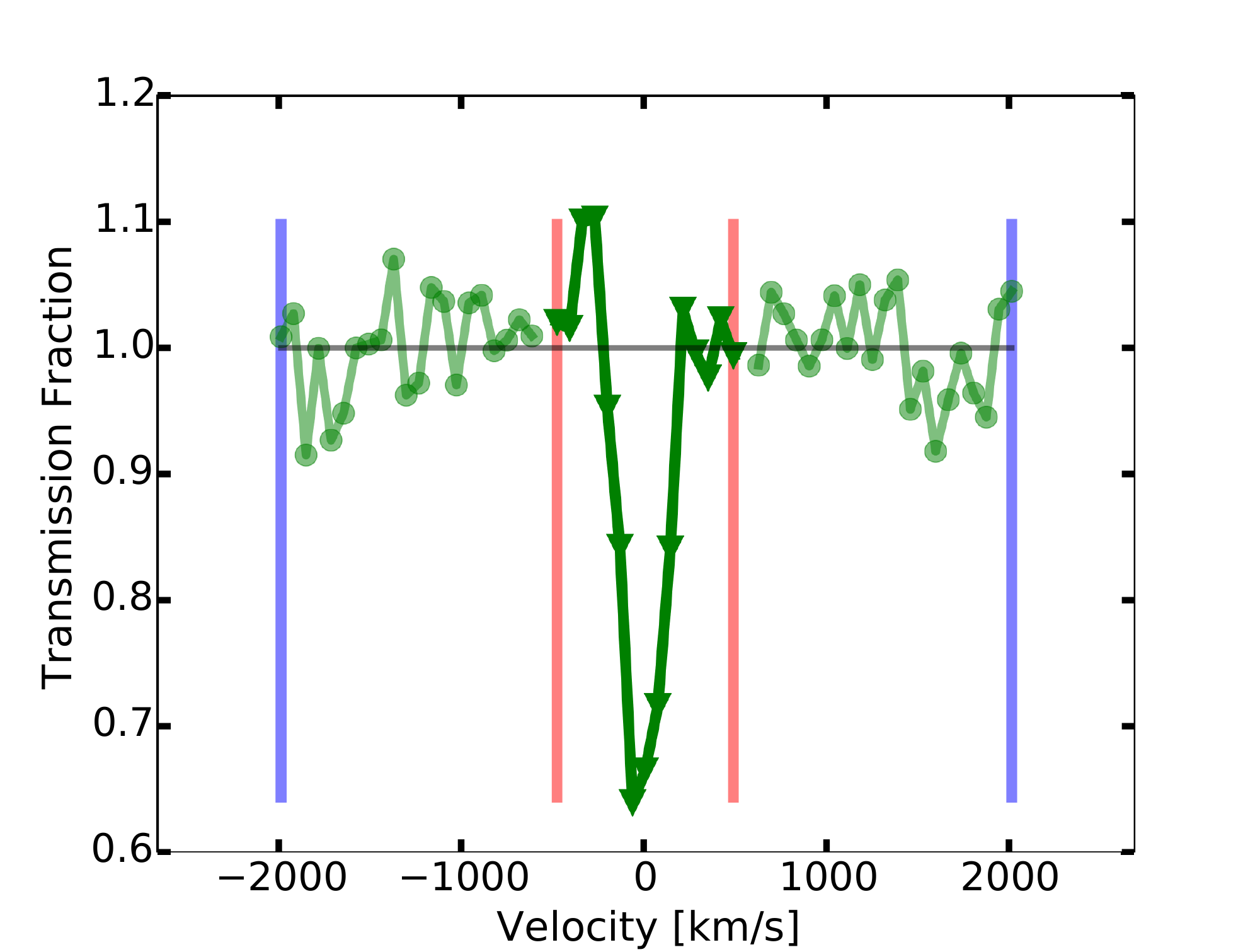}
\caption{ \label{fig:windows} The \emph{measurement} window (shown by
red vertical bars) and two \emph{continuum} windows (outer limits shown
by blue vertical bars) are indicated in this illustrating example of the
Al$\,$II$\,1670$ line of the DLA in the quasar spectrum MJD-plate-fiber
55182-3587-0100 in BOSS. The flux values in the \emph{continuum windows}
are fitted by a linear regression shown in magenta in the left panel.
The ratio of the flux to this linear regression gives the transmitted
fraction $F_i$ in the right panel.}
\end{figure}

  Figure \ref{fig:windows} illustrates the procedure in an example of an
Al\thinspace II 1670 absorption line. The red vertical bars show the
limits of the measurement window, and the blue bars are the outer limits
of the continuum windows. Green points show the values of the flux in
the left panel. The unweighted linear regression is the magenta line. The
right panel shows the result of dividing the flux values by this linear
regression, giving the transmitted fraction $F_i$.
The error on this transmitted fraction, $\sigma_{Fi}$,
is computed by dividing $\sigma_i$ by the same linear regression.

  Finally, we obtain the rest-frame equivalent width of a line $k$
simply by summing the absorbed fraction over all the $N_m$ pixels of the
measurement window,
\begin{equation}
\label{eq:W_raw}
W_k = \sum_{i=1}^{N_m} (1-F_i) \lambda_k\, p_0 ~, 
\end{equation} 
where $\lambda_k$ is the central rest-frame wavelength of the metal line
being measured, and $p_0=10^{-4}\times \log(10)=2.303\times 10^{-4}$ is
the width of the BOSS pixel in $\log\lambda$. Note that some pixels in
the measurement window may have $F_i>1$, and in fact, some equivalent
widths we compute are negative because of the noise. Still, as explained
before, they need to be included to have unbiased mean properties. The
error on this equivalent width is computed as
\begin{equation}
\label{eq:errW_raw}
\epsilon_k = \overline{\sigma}_F  N_m^{1/2} \lambda p_0 ~,
\end{equation} 
where $\overline{\sigma}_F^2$ is the average of the squared transmission
error, $\sigma_{Fi}^2=\sigma_i^2/C_i^2$, of the pixels in the two
continuum windows that are used to determine the continuum linear
regression. We use the noise in the continuum windows (instead of the
measurement window) to estimate the error $\epsilon_k$ because of the
correlation of $\sigma_{Fi}$ and $F_i$ discussed earlier, which can
induce an artificial correlation of the estimated value of $W$ and its
error. We note that the error $\epsilon_k$ accounts only for pixel
noise, assuming that it is uncorrelated in all the pixels, but excludes
any systematic errors due to the continuum fitting.

  We have tested our method by inserting mock lines in empty parts of a
spectrum and successfully recovering their equivalent widths within the
estimated uncertainty. We checked that equivalent widths estimated in
spectral intervals where we do not expect any absorbing line return
measurements consistent with zero.

\begin{figure*}
	\includegraphics[width=0.75\textwidth]{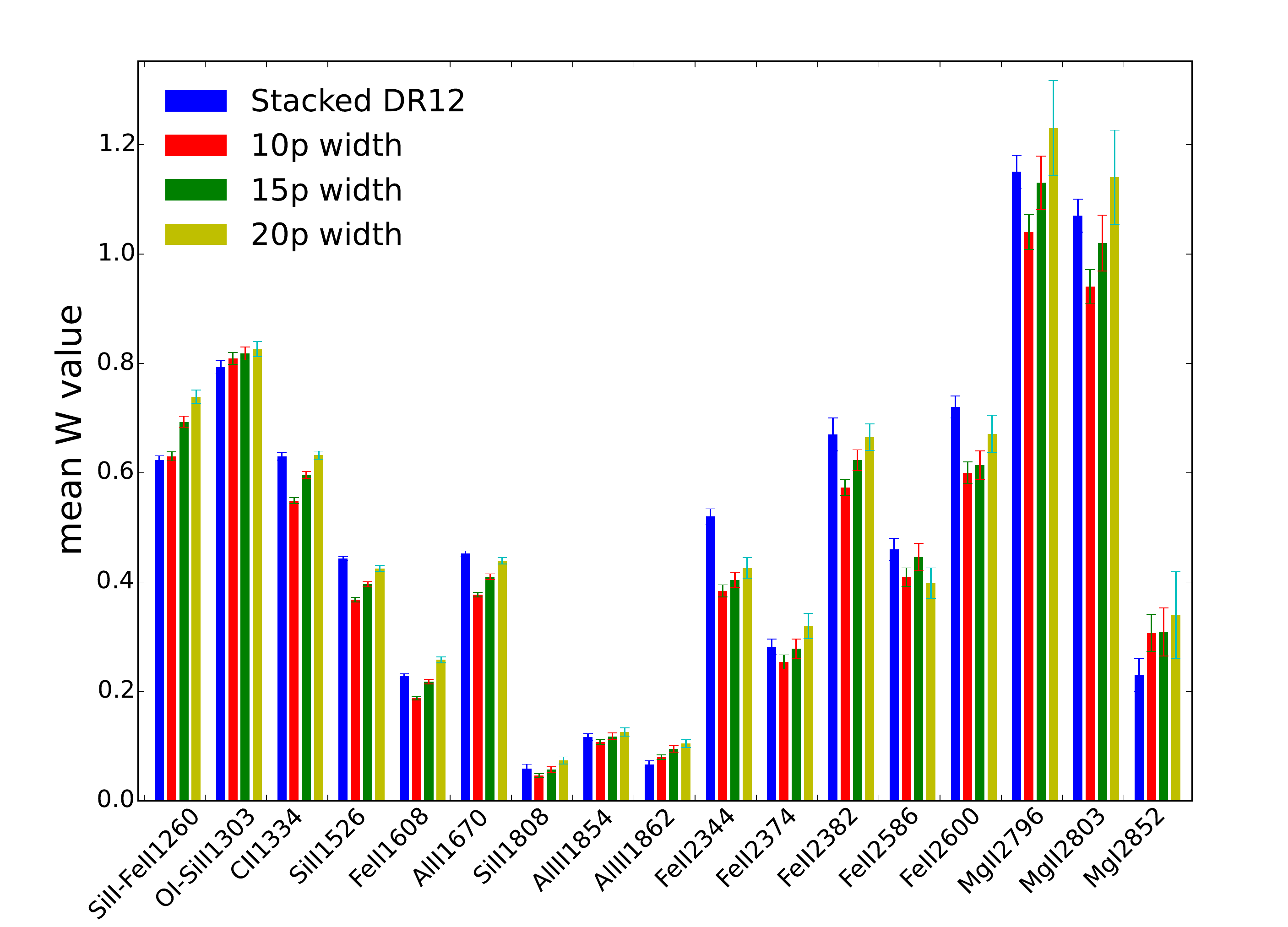}
	
	\caption{ \label{fig:mean_W_hist_win} Histogram comparing the stacked
	equivalent width from \protect\cite{lluis2014} with our mean equivalent width
	for the 17 lines used in this work, for three different sizes of the
	measurement window: 10, 15 and 20 pixels. Error bars show the standard
	deviation of the mean computed from the measured scatter among DLAs.}

\end{figure*}
 
  As a final test, in Figure \ref{fig:mean_W_hist_win} we compare the
mean equivalent width we derive for each of the 17 lines in Table
\ref{tab_lines} with the value obtained from the stacked absorption
spectrum presented in \cite{lluis2014}, and we also check the
sensitivity of this mean equivalent width to the size of our measurement
window by varying it from our standard width of 15 pixels to 10 and to
20 pixels. The mean values generally agree. The largest discrepancy
occurs for the Fe\thinspace II 2344 line, for which our method yields an
equivalent width $\sim 20$\% smaller than in \cite{lluis2014}. Our
derived $\overline W$ generally increases with the measurement window
width, as expected because of the effect of redshift errors. The overall
agreement is relatively good for a 15 pixel width, which we choose as
the optimal and standard one from this plot.

\section{Definition of the Metal Strength}
\label{sec:sel_par}

 ~\par We now define a {\it metal strength} for each DLA intended to
provide a weighted average of the strength of the observable associated
metal lines with an optimal signal-to-noise, which can be used in
spectra with high noise where individual lines are generally barely
detectable. If the metal lines that have been measured for a given DLA
are labeled by the index $k$, with equivalent widths $W_k$ and error
$\epsilon_k$ from equations (\ref{eq:W_raw}) and (\ref{eq:errW_raw}),
the metal strength is defined as
\begin{equation}
\label{eq:sel_par}
S = \frac{ \sum_k \left( \overline{W}_k /\epsilon_k \right)^2 \cdot
 \left(  W_k /\overline{W}_k \right)  }{ \sum_k \left(
  \overline{W}_k /\epsilon_k \right) ^2 } ~,
\end{equation}
where $\overline{W}_k$ is the mean equivalent width of line $k$ over
all the DLAs for which it is measured. The
metal strength is equal to unity if all the lines of a DLA have the
mean value $\overline W_k$, and is in general a measure of the ratio
of the equivalent widths to their mean value. Each line is weighted by
$(\overline W_k/\epsilon_k)^2$, which is the expected squared
signal-to-noise ratio of the measurement if the line has an equivalent
width equal to the mean. These mean values are listed in the
fourth column of Table \ref{tab_lines}. The error of this metal strength
is computed from the errors of each individual line as
\begin{equation}
\label{eq:err_sel}
\epsilon_S = \left[ \sum_k \left(
 \frac{\overline{W}_k }{\epsilon_k} \right) ^2 \right] ^{-1/2} ~. 
\end{equation}

  This metal strength parameter reflects how strong the metal lines of a
DLA are, and is chosen as a quantity that can be measured with an
optimal signal-to-noise ratio from spectra similar to those in the BOSS
survey. The metal strength is expected to increase both with the metal
abundances and the velocity dispersion. A larger
velocity dispersion implies a wider line which is less saturated,
thereby yielding a larger equivalent width for a fixed column density
of metals.

\begin{figure}
    \centering
    \includegraphics[width=\columnwidth]{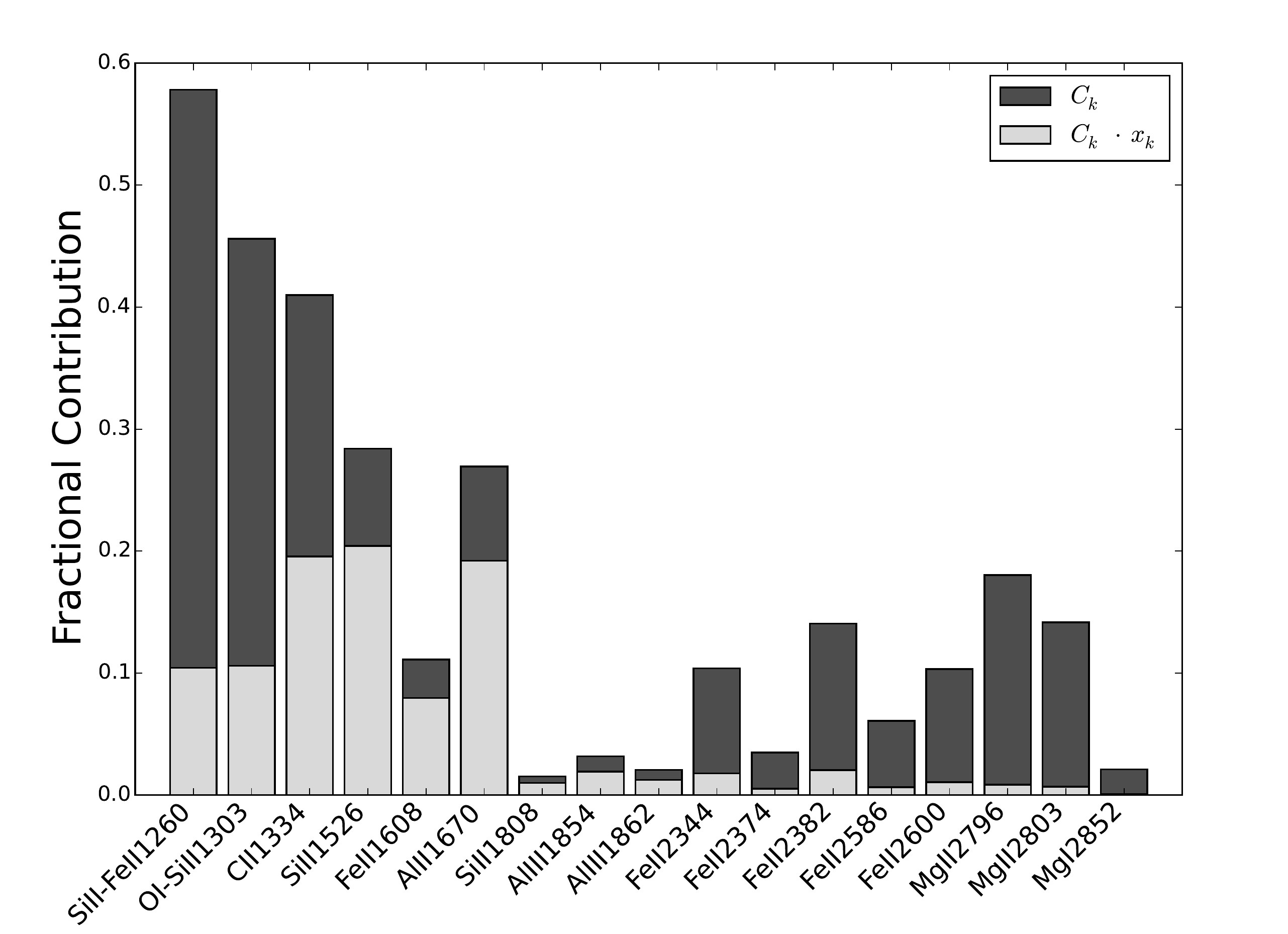}
    \caption{\label{fig:contribution} {\it Dark grey histogram}: mean
contribution $C_k$ of each line to the metal strength of DLAs in which the
line is measured. {\it Light grey histogram}: mean contribution to all
the DLA sample, equal to $C_k x_k$, where $x_k$ is the fraction of DLAs
in which the line is measured. The sum of all $C_k x_k$ adds to unity.}
    
\end{figure}

  To visualize the metal lines that are most important to determine $S$,
we define the mean contribution $C_k$ of metal line $k$ as follows:
\begin{equation}
\label{eq:cont}
C_k =  \frac{1}{N_k}\, \sum_{i=0}^{N_k} 
\frac{\left( \overline{W}_k /\epsilon_{k,i} \right)^2}{
 \sum_j \left( {\overline{W}_j} /\epsilon_{j,i} \right)^2 } ~ ,
\end{equation}
where the sum over $i$ includes all the $N_k$ DLAs in which the metal
line $k$ has been measured, $\epsilon_{j,i}$ is the error of the metal
line $j$ in the DLA $i$, and the sum in the denominator adds the weights
of all metal lines $j$ that have been measured in the DLA $i$. This
means that $C_k$ is the mean contribution to the metal strength $S$ from
line $k$, restricted only to the fraction $x_k$ of DLAs in which this
metal line has been measured. The average contribution to all the DLA
sample is the product $C_k x_k$, which adds to unity when summed over
the 17 metal lines. These quantities are listed in Table
\ref{tab_lines}, and shown also in Figure \ref{fig:contribution}.

 The lines that are important for most DLAs, with the highest value of
$C_k x_k$, are the first 6 at short wavelength. A few of the longer
wavelength ones are important only in a small fraction of DLAs.
The fraction $x_k$ is small for the first line, the SiII-FeII blend,
which has the shortest wavelength, because any lines that are close to
the quasar \lya emission line or that fall in the \lya forest region are
excluded. Then, $x_k$ increases with increasing line wavelength until
it starts decreasing rapidly above $\sim 2000\angs$ due to the exclusion
of metal lines when a sky line falls within the measurement window, and
the large abundance of sky lines in the red part of the spectrum.

\begin{figure}
   \centering
   \includegraphics[width=\columnwidth]{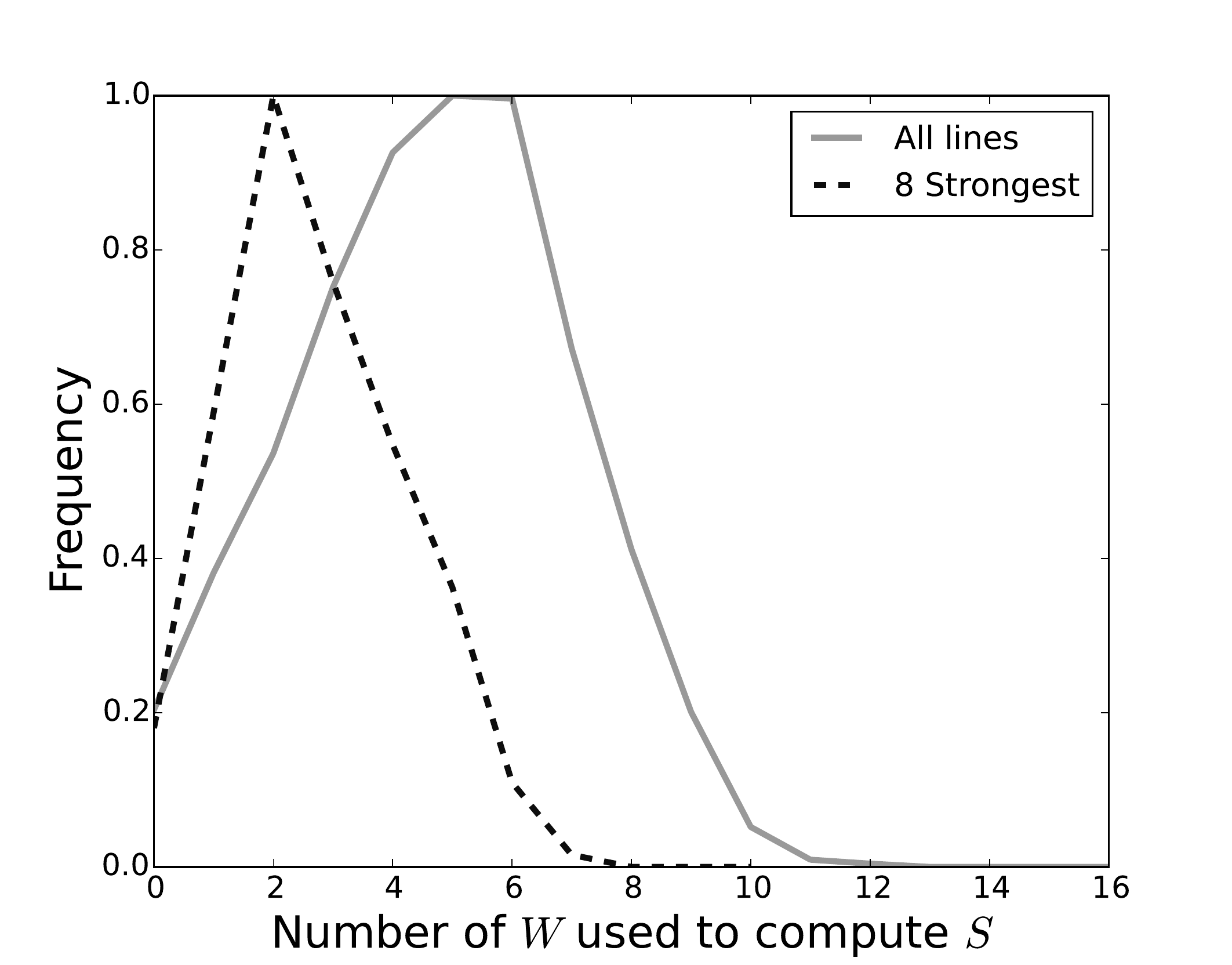}
   \caption{   \label{fig:n_lin} Distribution of the number of lines
measured for a DLA among all 17 lines in Table \ref{tab_lines}
(gray solid line), and among the 8 lines with a mean contribution
$C_k > 0.15$ (defined below in eq. \ref{eq:cont}).}

\end{figure}

  The distribution of the number of lines contributing to the
determination of $S$ in equation \ref{eq:sel_par} (which satisfy the
criteria specified in \S \ref{subs:select_met}) is shown in Figure \ref{fig:n_lin}
as the grey solid line. The average number of contributing lines is
between 6 and 7. The black dashed line is the distribution of
the number of measured lines among only the 8 lines with $C_k > 0.15$ in
Table \ref{tab_lines}. In most cases, there are only 2 or 3 of these
highly contributing lines.

\subsection{Correction for the dependence of $\overline W_k$ on $N_{\rm HI}$.}
\label{subs:corr}

  We expect the mean equivalent width of any metal line to increase with
$N_{\rm HI}$, because for a fixed metal abundance and internal velocity
distribution, the metal column density should be proportional to the
hydrogen column density. This dependence was confirmed in
\cite{lluis2014} from stacked spectra of DLAs in different bins of
$N_{\rm HI}$ (see their Table 3). Our goal is to classify DLAs
depending on a parameter that reflects only their metal content and
velocity dispersion (which affects the degree of saturation of the
line), but not on $N_{\rm HI}$.

  For this purpose, we define the HI-corrected metal strength,
$S_{\rm HI}$, in the same way as $S$ in equation \ref{eq:sel_par}, but
replacing the constant values of $\overline W_k$ by mean equivalent
widths that depend linearly on $\log(N_{\rm HI})$:
\begin{equation} \label{eq:lin_Hd}
 \overline{W}_k(N_{\rm HI}) = a_k (\log(N_{\rm HI}) - 20) + b_k ~.
\end{equation} 
We fit the parameters $a_k$ and $b_k$ by dividing the DLA sample into
the following 5 intervals of $\log(N_{\rm HI})$: [20.0, 20.20),
[20.2, 20.4), [20.4, 20.63), [20.63, 21.0), and [21.0, 22.2]. Then, for
each metal line, we compute the mean value of $W_k$ and
$\log(N_{\rm HI})$ in the five intervals, and we make an unweighted
linear regression of these 5 points.

 \begin{table}

\centering
\caption{ \label{tab_reg} Values for the linear regression fitted to the
dependence of $W$ on $\log(N_{\rm{HI}})$ in equation \ref{eq:lin_Hd}.
The first two columns use values of $W$ from the \textit{stacking} technique
shown in \protect\cite{lluis2014}, and the latter 2 columns use $W$ measured
with our method, which are used for the analysis in this paper.
}
	\vspace{3 mm}
	
	\begin{tabular}{ccccc}

	\hline
	Name  &   $a_k$ \textit{stacking} & $b_k$ \textit{stacking} &   $a_k$  & $b_k$ \\
	\hline
	\hline
	SiII-FeII$\,1260 	$  &		0.316	&		0.439	&  0.267 &  0.568   \\
	OI-SiII$\,1303 	$  &		0.591	&		0.482	&  0.651 &  0.489   \\
	CII$\,1334  	$  &		0.311		     &		0.443	&  0.391 &  0.403   \\
	SiII$\,1526  	$  &		0.327		&		0.270	&  0.324 &  0.231    \\
	FeII$\,1608 	$  &		0.263		&		0.089	&  0.287 &  0.0744  \\
	AlII$\,1670  	$  &		0.338		&		0.273	&  0.368 &  0.233    \\
	SiII$\,1808  	$  &		0.101		&		0.0041	&  0.110 &  0.00105   \\
	AlIII$\,1854 	$  &		0.091		&		0.068	&  0.147 &  0.0478    \\
	AlIII$\,1862 	$  &		0.052		&		0.038	&  0.0399 &  0.0759  \\
	FeII$\,2344  	$  &		0.471		&		0.263	&  0.568 &  0.123     \\
	FeII$\,2374  	$  &		0.374		&		0.085	&  0.337 &  0.104   \\
	FeII$\,2383  	$  &		0.559		&		0.370	&  0.513 &  0.360    \\
	FeII$\,2587  	$  &		0.531		&		0.179	&  0.574 &  0.122   \\
	FeII$\,2600  	$  &		0.491		&		0.455	&  0.528 &  0.304     \\
	MgII$\,2796 	$  &		0.595  		&		0.841	& 1.008 &  0.583     \\
	MgII$\,2803 	$  &		0.650		&		0.731	&  0.923 &  0.576    \\
	MgI$\,2853  	$  &		0.217	  &		0.092	& 0.208 &  0.135   \\

	\hline	
	
	\end{tabular}
	
\end{table}

  The result is shown in Table \ref{tab_reg}, in columns 4 and 5. These
linear regressions to the mean equivalent width as a function of
$N_{\rm HI}$ were also obtained in \cite{lluis2014} from stacked
spectra, and their results are shown in columns 2 and 3. The values are
in general similar, and differences can be attributed to the different
methods of determining the quasar continuum and weighting the DLAs.
 
  The mean equivalent widths calculated from equation \ref{eq:lin_Hd}
for the column density of each DLA are then used to compute the
corrected metal strength $S_{\rm HI}$ with the same equation
\ref{eq:sel_par} as for $S$.

\section{Results }
\label{sec:results}

  ~\par We now present the distribution of the metal strength parameter $S$
and its variation with column density and redshift. We also examine if
the metal strength distribution has any dependence on the
continuum-to-noise ratio in the \lya forest, and the computed error
$\epsilon_S$, as a test of any possible variation of the purity of the
DLA catalogue that might affect derived statistical properties of DLAs as
a function of $S$. As the main product of this work, we present the
catalogue of metal line equivalent widths and the metal strength parameter
with its error, both corrected and uncorrected for its $N_{\rm HI}$
dependence, for all the DLAs in DR12-DLA.

 
\subsection{The metal strength distribution}
\label{subs:win_dist}

\begin{figure}
    
    \centering
   \includegraphics[width=0.95\columnwidth]{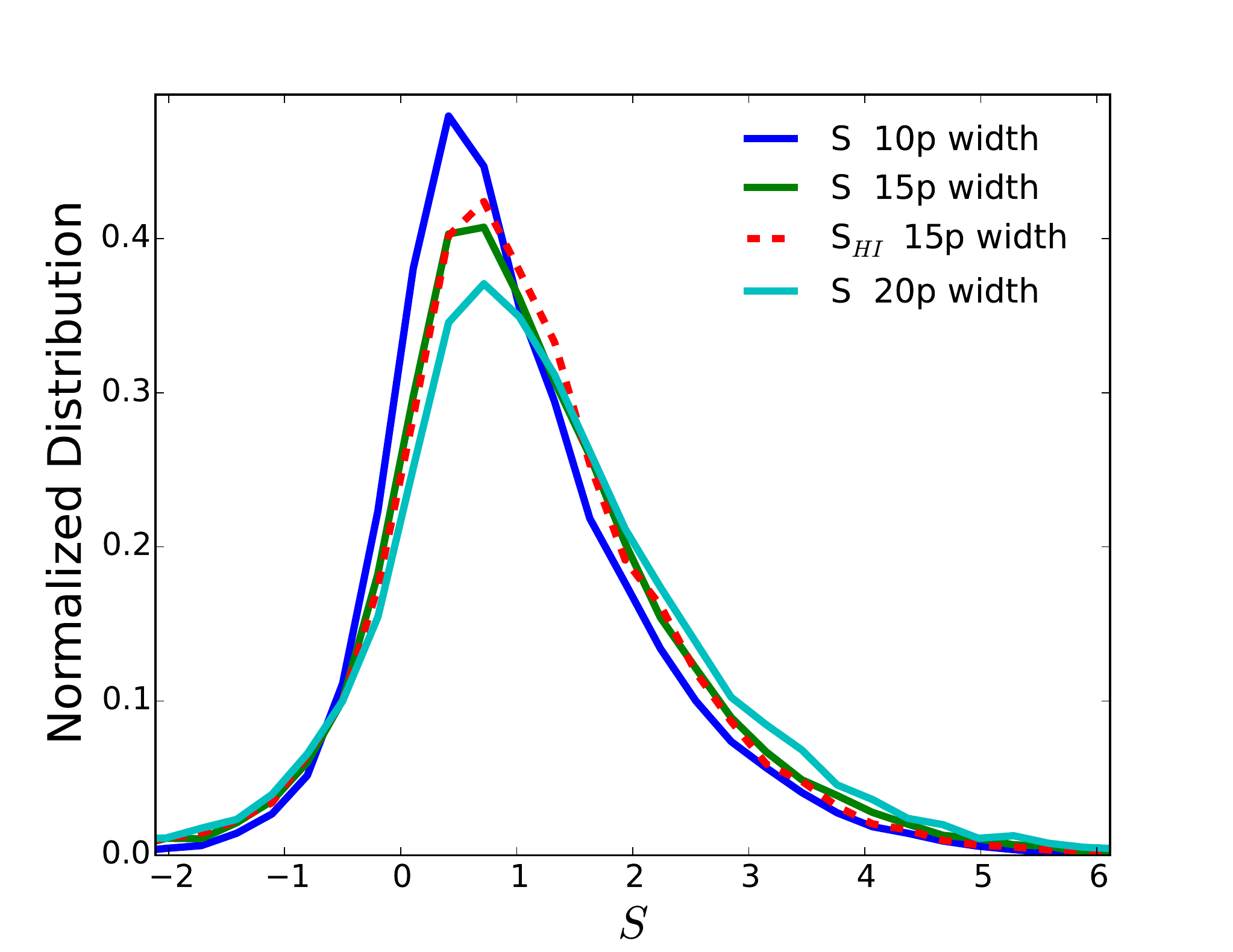}    
	\includegraphics[width=0.95\columnwidth]{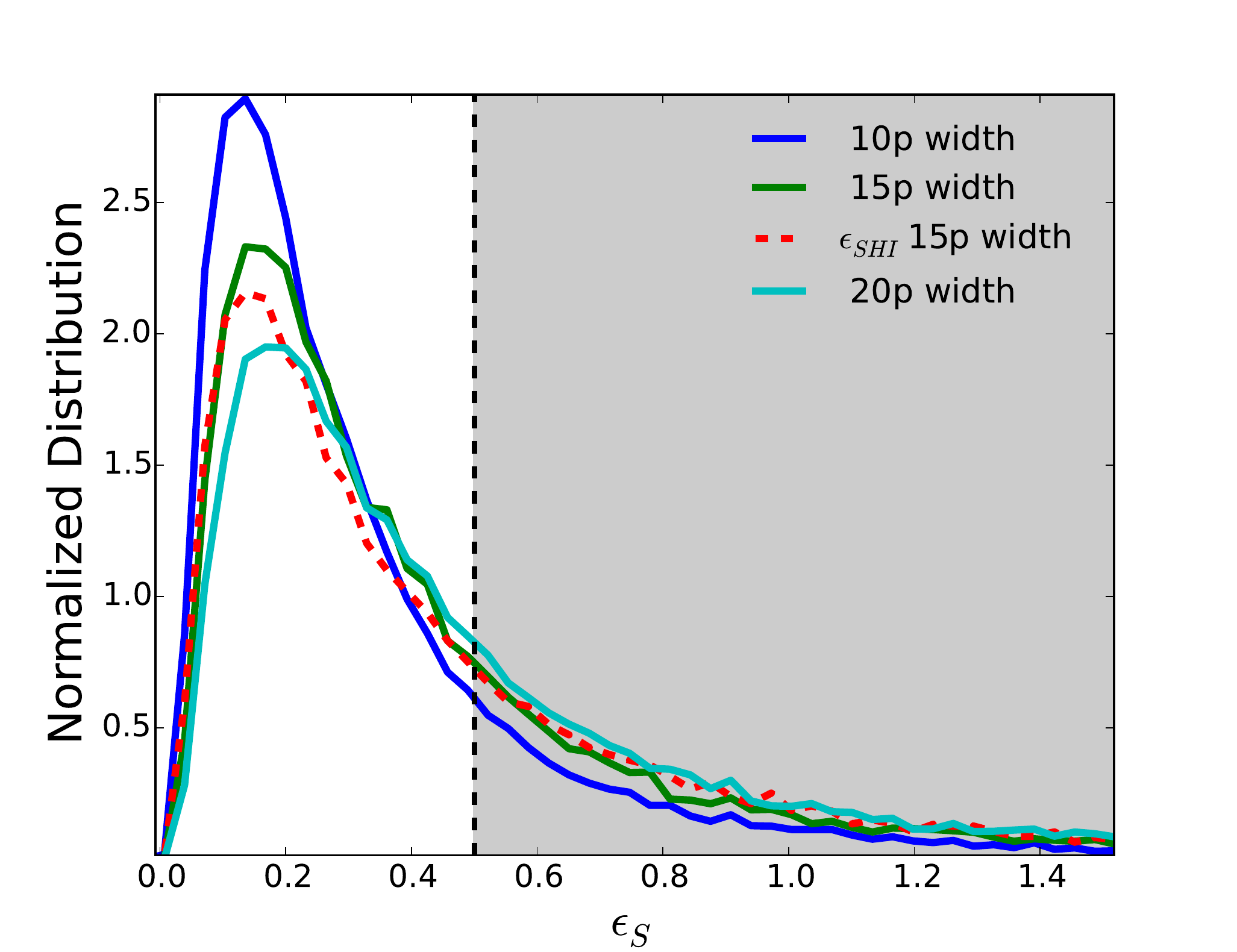}
	
	\caption{\label{fig:win_distributions}
{\it Top panel:} Metal strength distribution for three different widths
of the measurement window (10, 15 and 20 pixels), compared also with
the distribution of $S_{\rm HI}$ for our standard 15 pixel window.
Only DLAs with error $\epsilon_S < 0.5$ are included.
{\it Bottom panel:} Distribution of the error $\epsilon_S$ for the
same three measurement window widths, and the $N_{\rm HI}$-corrected
version for the 15 pixel window.
}
\end{figure}

  The distribution of the metal strength $S$ is shown in the top panel
of Figure \ref{fig:win_distributions}, including only DLAs with an
error $\epsilon_S < 0.5$. In general, we exclude DLAs with larger error
to better see their intrinsic properties. The distribution we measure is
the convolution of the true distribution with the error
distribution. Systems with negative $S$ should be a consequence of the
error, and this is consistent with the shape of the curves at $S<0$,
except at $S < -1$ where the extended tail indicates the presence of
non-Gaussian errors that are likely due to systematics in the continuum
fitting. The $S$ distribution becomes wider with the measurement window
width owing to increased errors. The $N_{\rm HI}$ correction practically
does not affect this distribution.

  The bottom panel shows the distribution of $\epsilon_S$. The error
increases with the measurement window simply because of the increased
number of pixels in equation \ref{eq:errW_raw}. The majority of DLAs
have errors $\epsilon_S<0.5$, and selecting this subset helps make the
distribution of $S$ we obtain in the top panel less distorted from the
true one. We see that this subset of DLAs have an error $\epsilon_S$
smaller than the true dispersion in $S$, and therefore can be divided
into smaller subsets corresponding to intervals in $S$ to measure other
DLA properties as a function of $S$.
In future papers, we will study the dependence of the DLA bias factor
and the mean absorption spectrum on the parameter $S$. 

  Results in the rest of the paper are shown only for DLAs with
$\epsilon_S < 0.5$, unless otherwise specified.

\subsection{Metal strength dependence on $N_{\rm HI}$ and redshift}
\label{subs:Hd_dep}

\begin{figure}
    \centering  
	\includegraphics[width=\columnwidth]{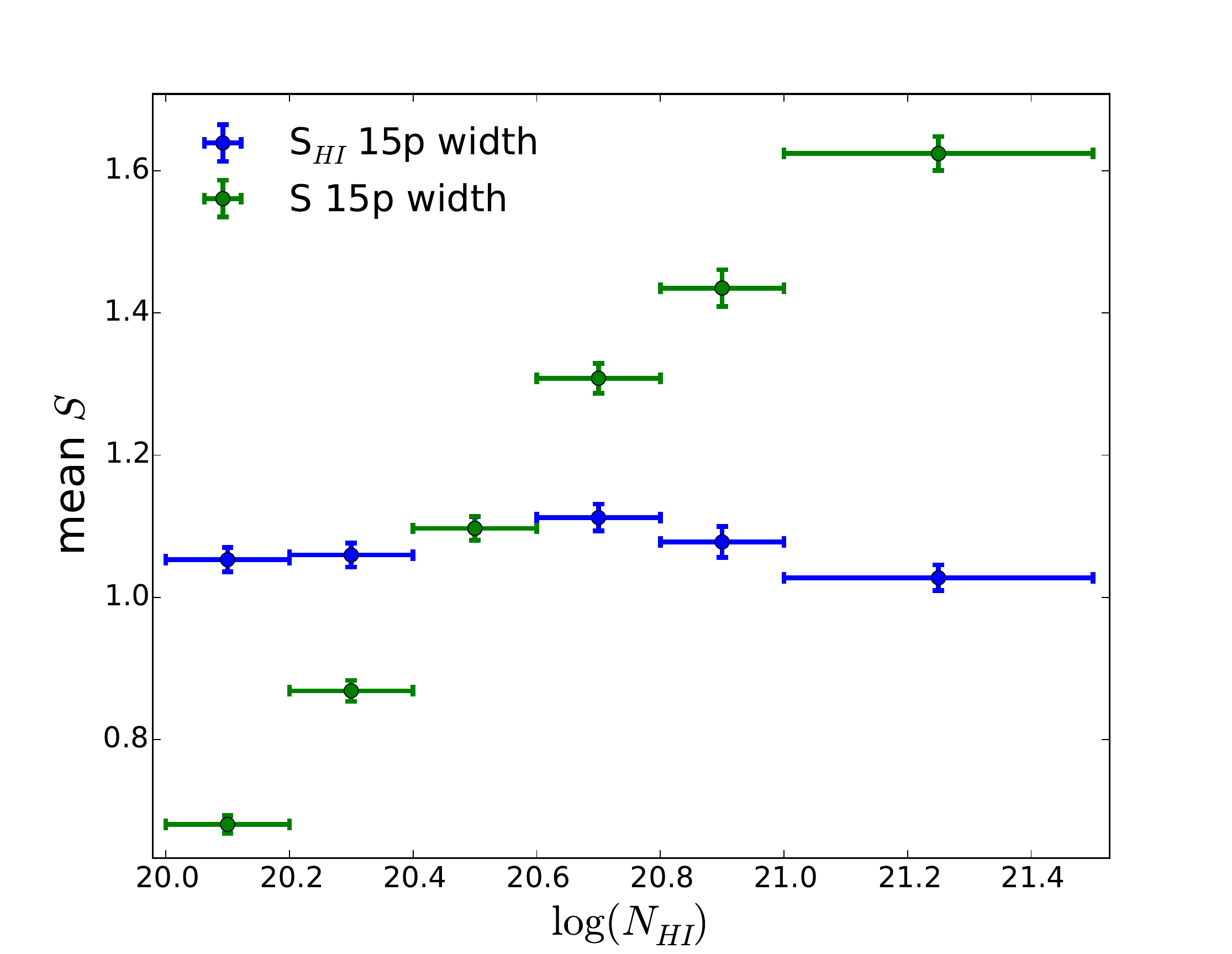}
	
	\caption{\label{fig:S-Hd_dependence} \textit{Green points} show
the mean metal strength at different column density intervals indicated
by the error bars.
\textit{Blue points} show the mean of the corrected metal strength
$S_{\rm HI}$, as defined in \ref{subs:corr}. 
 }
 \end{figure}
 
  Figure \ref{fig:S-Hd_dependence} shows the mean value of $S$ in
several column density intervals as the green points. As expected, $S$
increases with $N_{\rm HI}$ simply due to the increasing metal column
densities. After applying the correction discussed in \S \ref{subs:corr}
(blue points), the mean value of $S_{\rm HI}$ becomes indeed nearly
constant. Note that the mean value of $S$ should be unity for the whole
DLA sample; when we eliminate DLAs with $\epsilon_S > 0.5$, the mean
value of $S$ increases slightly above unity because of a decrease of the
mean $S$ with $\epsilon_S$ that will be discussed below.
 
\begin{figure}
    \centering
      \includegraphics[width=\columnwidth]{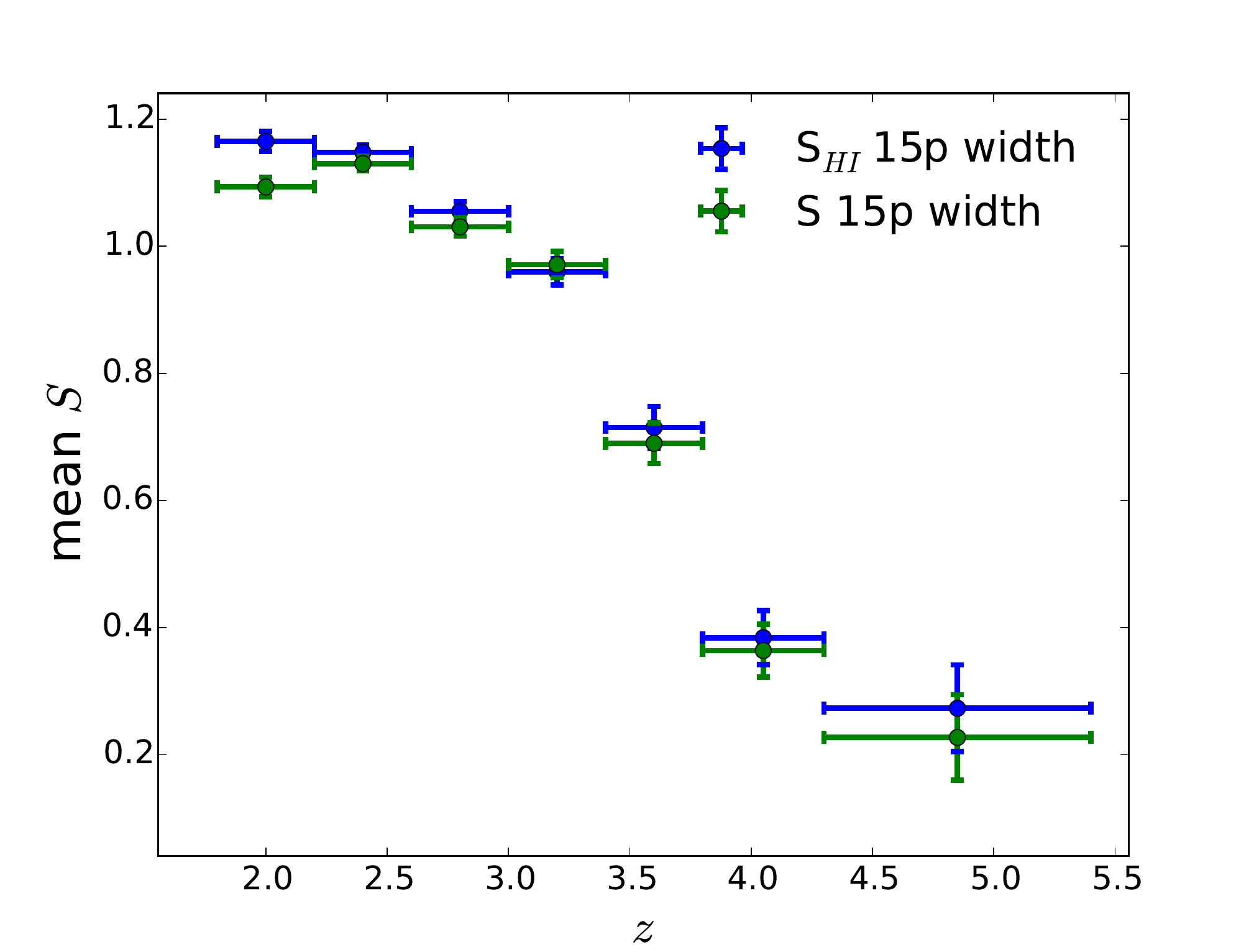}  
    \caption{\label{fig:S_evol_Z} Evolution of the mean metal strength
 $S$ (green points) and $S_{\rm HI}$ (blue points) with redshift.}
\end{figure}

  The evolution of the mean metal strength with redshift is shown in
Figure \ref{fig:S_evol_Z}. The mean $S$ is roughly constant at $z<3.4$,
and then drops rapidly at higher redshift. Some of this decline should
be due to the well-known decrease of the average DLA metallicity with
redshift (e.g., \cite{Kulkarni2005,Rafelski2012}). A decreasing velocity
dispersion with redshift may also contribute because the mean value of
$S$ increases with velocity dispersion at a fixed metal column density
owing to the effects of line saturation. However, the dependence of the
purity of the catalogue on redshift may be the main effect causing the
rapid drop at $z>3.4$. Only $\sim$ 10\% of our DLA sample is at $z>3.5$
(see Fig. 1 of \cite{lluis2014}), and the difficulty in detecting DLAs
is greatly increased at high redshift because of the increased mean
absorption of the \lya forest, which is likely decreasing the purity of
the catalogue. More complete studies revising the DLA detection method
will be needed to resolve this issue.

\begin{figure}
    \centering
   \includegraphics[width=0.95\columnwidth]{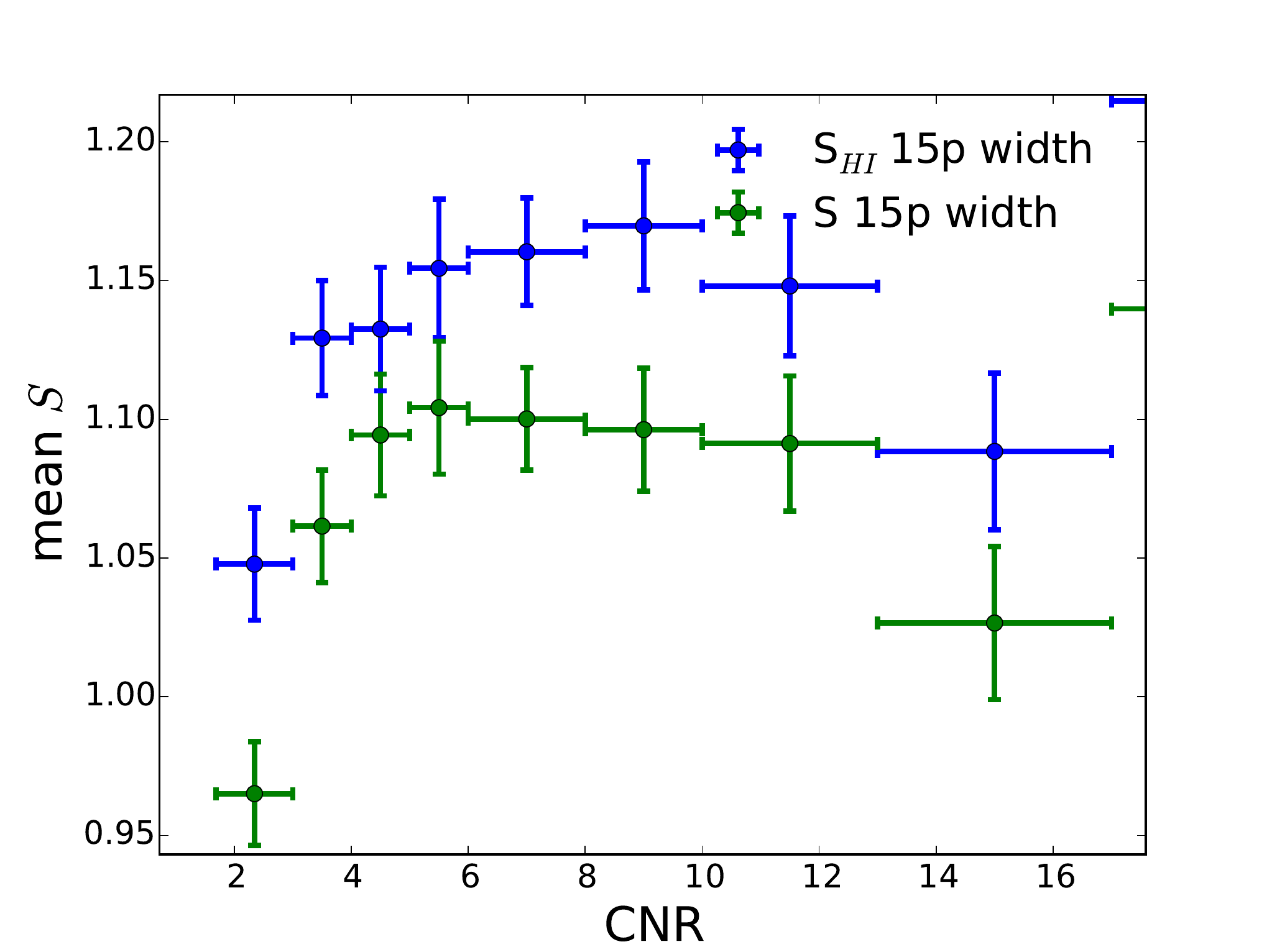}    
	\includegraphics[width=0.95\columnwidth]{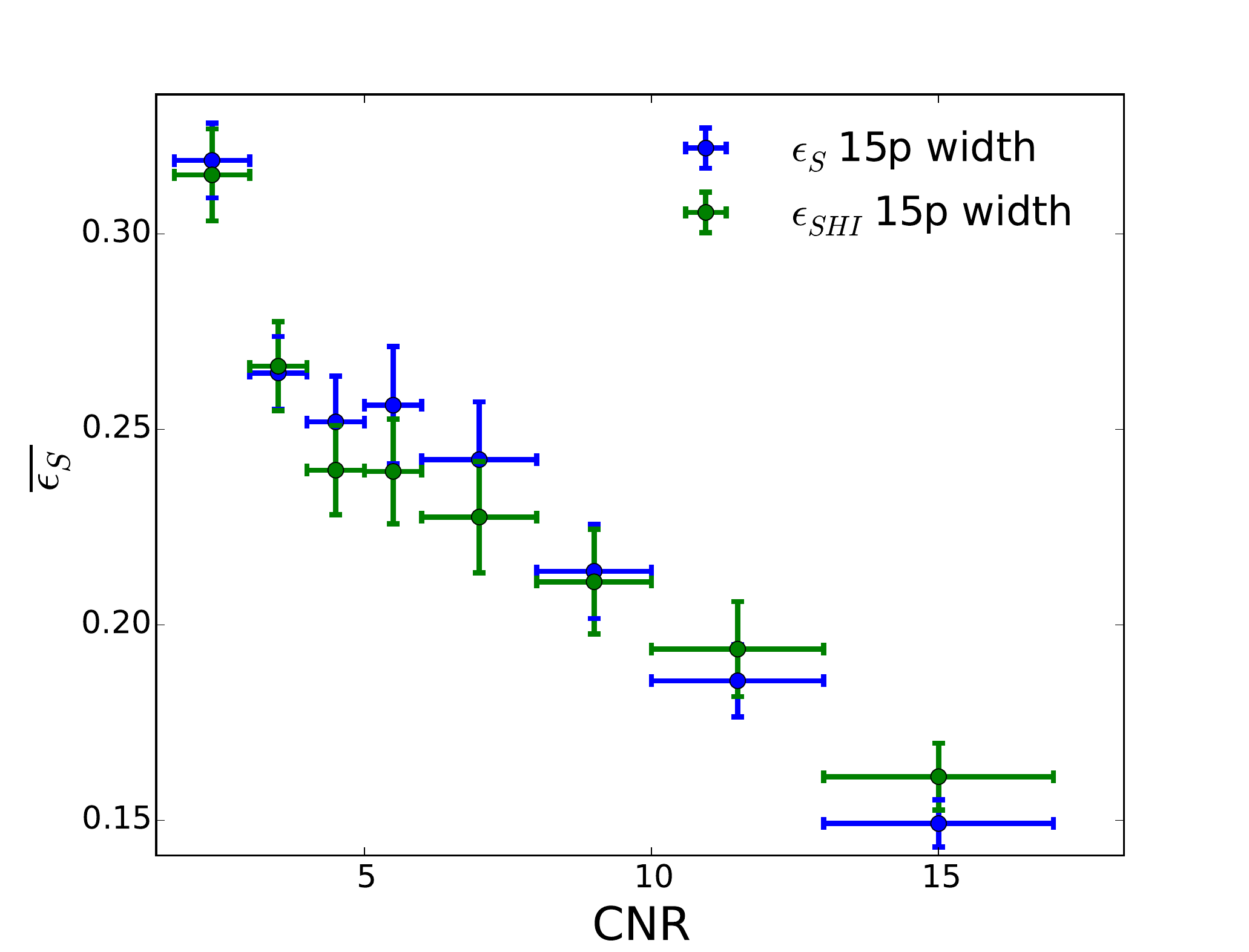}
	
\caption{\label{fig:S-CNR}\label{fig:S_evol_CNR}
{\it Top panel:} Mean metal strength $S$ as a function of the
continuum-to-noise ratio in the \lya forest region. {\it Bottom panel}:
Mean error $\epsilon_S$ as a function of continuum-to-noise ratio.
}
\end{figure}

\subsection{Dependence of the $S$ distribution on $\epsilon_S$ and the
purity of the DLA catalogue}
\label{sub:S_evol_CNR}

\begin{figure}
    \centering
   \includegraphics[width=0.95\columnwidth]{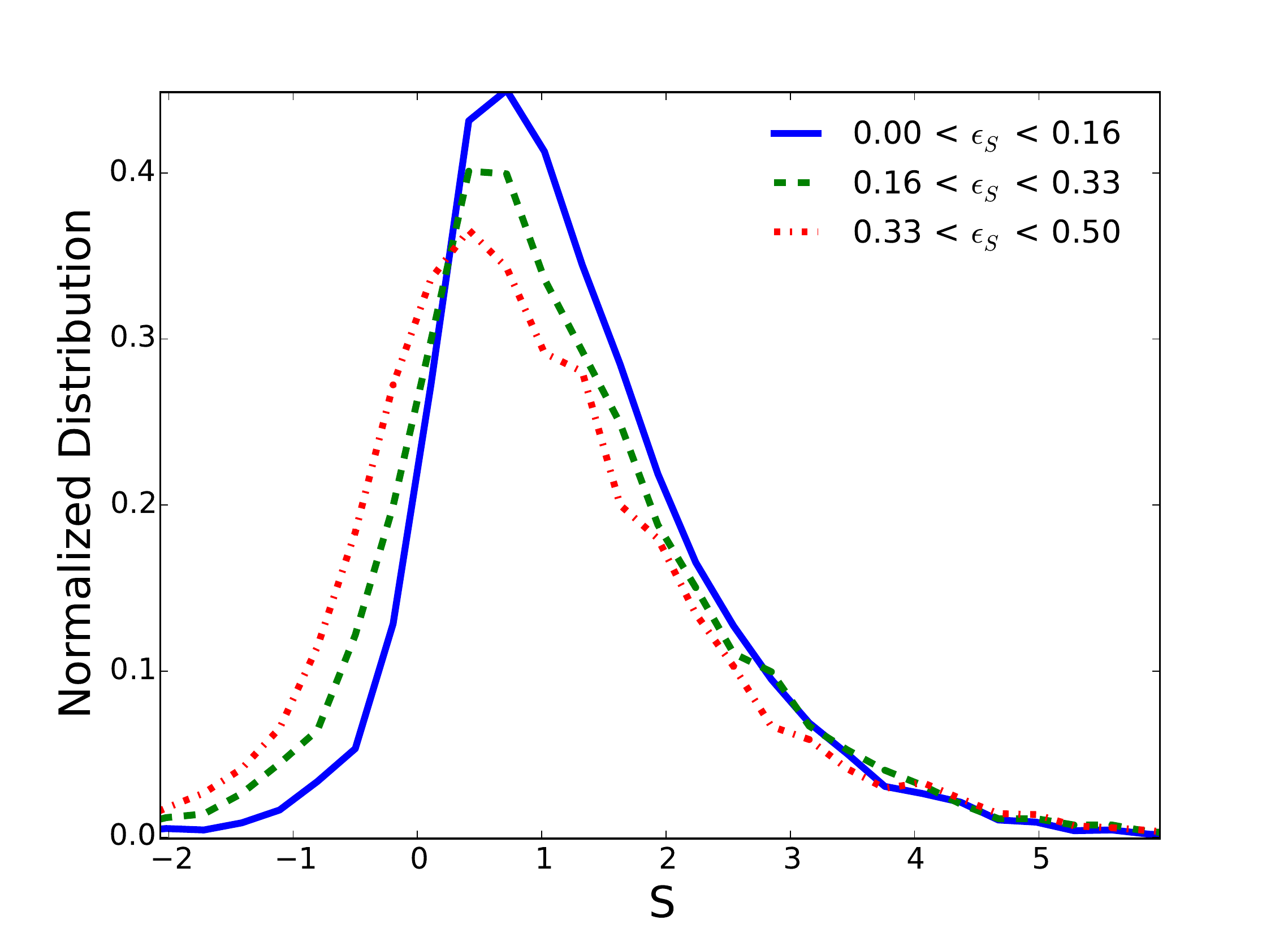}    
	\includegraphics[width=0.95\columnwidth]{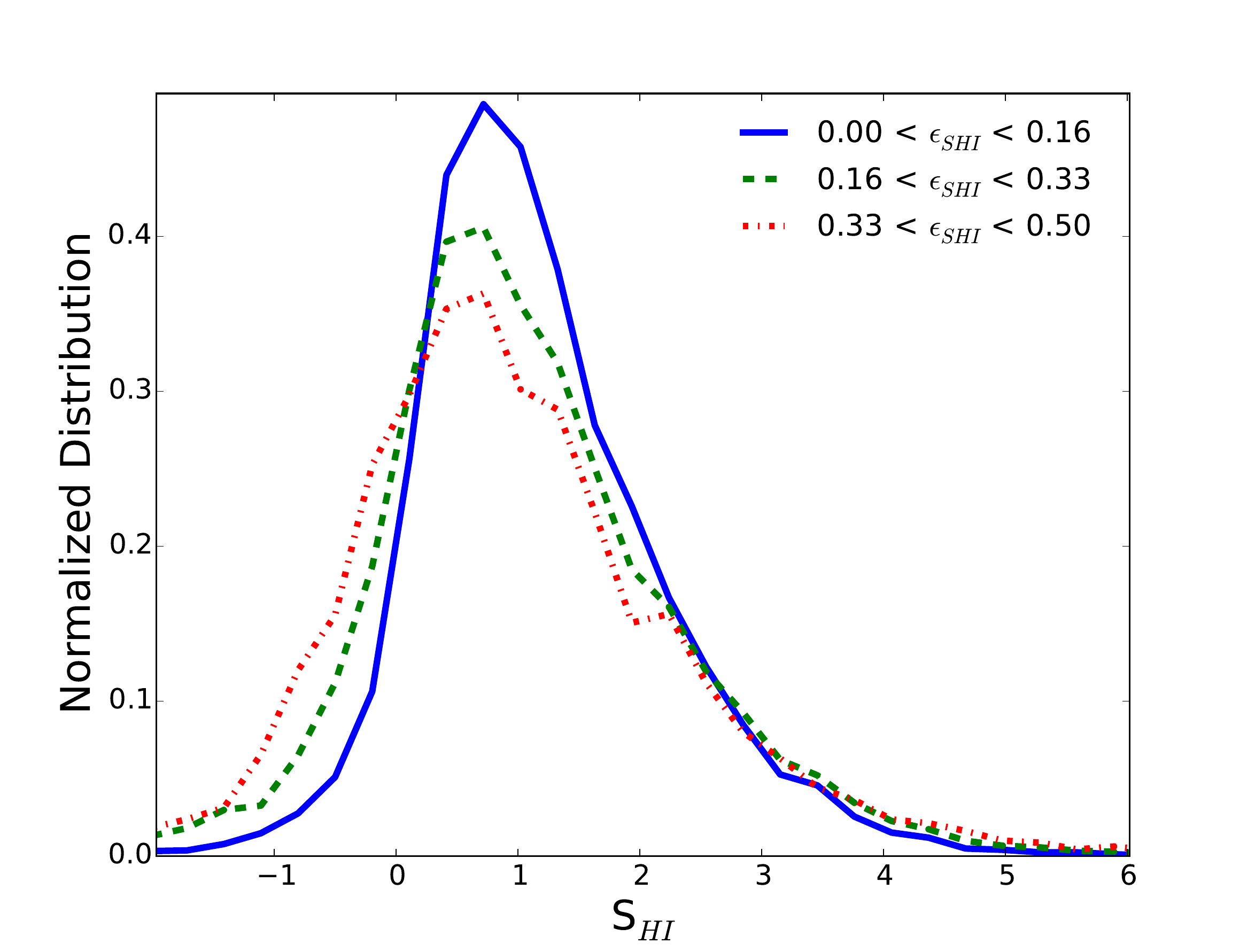}
	
	\caption{\label{fig:S-Serr_distribution}
\textit{Upper panel}: Distribution of the DLA metal strength for 3
intervals of the error $\epsilon_S$.
\textit{Lower panel}: Same for the corrected metal strength $S_{\rm HI}$. 
 }
 \end{figure}

  Our main goal in defining and measuring the metal strength $S$ for
individual DLAs is to be able to measure average quantities of the DLAs
as a function of $S$. An important difficulty to accomplish this is the
imperfect purity of the catalogue. For example, false DLAs arising purely
from spectral noise should have a mean value of $S$ and bias factor
equal to zero, and this can induce a spurious increase of the bias factor
with $S$. Other false DLAs may be dense \lya forest regions with a
column density and mean $S$ much lower than DLAs, and a different bias
factor.

\begin{figure}
    \centering
   \includegraphics[width=0.88\columnwidth]{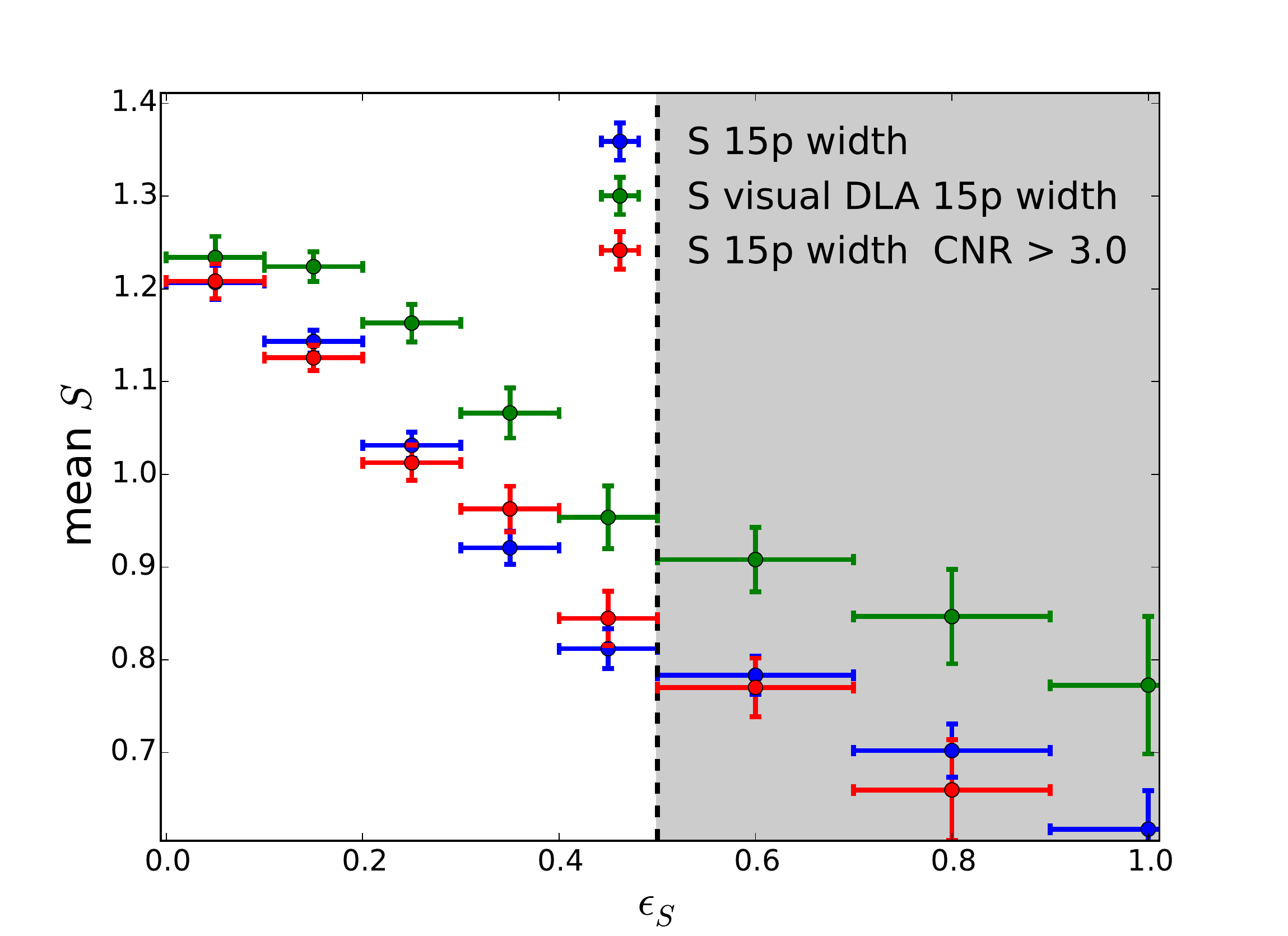}    
	\includegraphics[width=0.88\columnwidth]{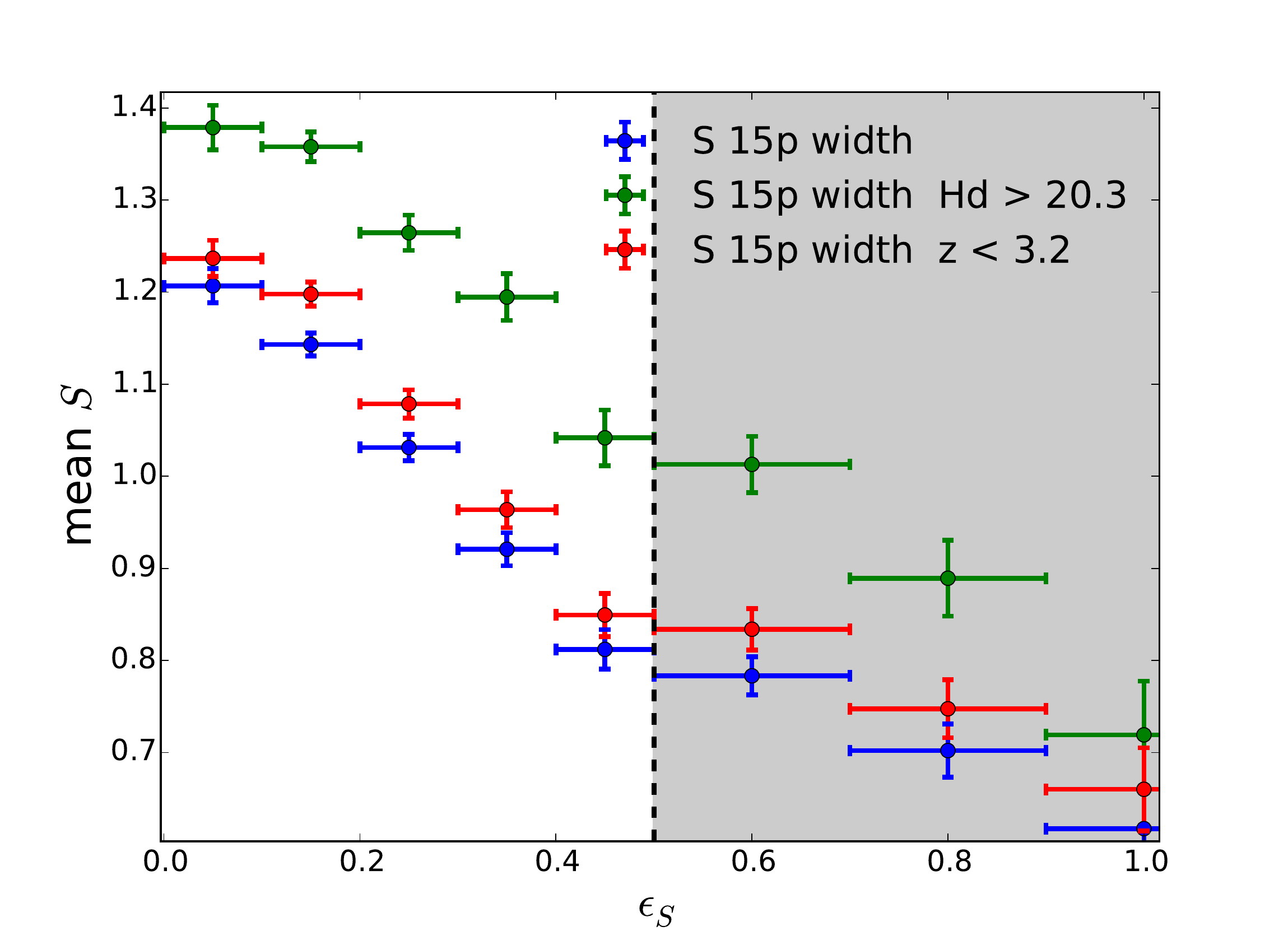}
	\includegraphics[width=0.88\columnwidth]{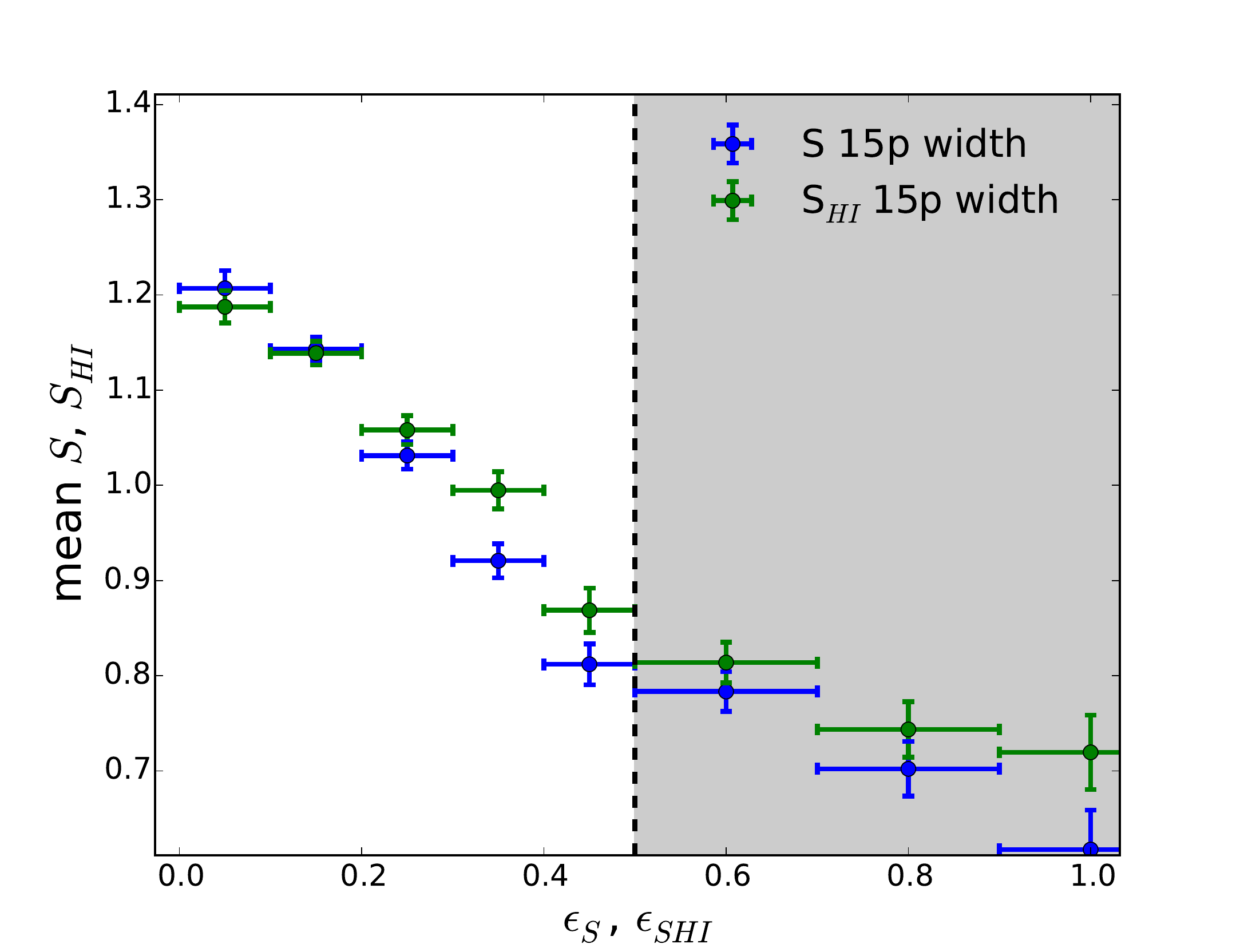}
	\caption{\label{fig:S-Serr-dependences}
 Mean metal strength as a function of the error $\epsilon_S$ for several
sub-samples. \textit{Upper panel}: standard $\epsilon_S<0.5$ sample
({\it blue}), compared to the sub-sample of DLAs with ${\rm CNR} > 3$
({\it red}) and to the visually identified DLAs ({\it green}).
\emph{Middle panel}: standard sample compared to DLAs with
$N_{\rm HI} > 10^{20.3} {\rm cm}^{-2}$ ({\it green}) and to DLAs with
$z< 3.2$ ({\it red}). \emph{Lower panel}: standard sample compared both
for the uncorrected (blue) and corrected (green) metal strength.
}
 \end{figure}
 
  A first test we can do to investigate possible variations of the
purity with $S$ is to check if the mean value of $S$ depends on the
continuum-to-noise (CNR) ratio of the \lya forest region, where the
DLA absorbers are searched. We expect the purity of the catalogue to
increase with CNR. This is shown in the top panel of Figure
\ref{fig:S-CNR}, both for the uncorrected and $N_{\rm HI}$-corrected
cases. The mean $S$ is consistent with a constant value for
$CNR > 3$, suggesting that as long as we restrict any analysis to this
minimum CNR, there should not be strong effects of the catalogue impurity
caused by spectral noise. The bottom panel shows that, as expected, the
error $\epsilon_S$ decreases with CNR (since they both reflect the
noise amplitude in the spectrum, although on different wavelength
ranges). In both the uncorrected and corrected cases, the sample is
restricted to DLAs with $\epsilon_S < 0.5$, or $\epsilon_{\rm SHI} < 0.5$.

\begin{figure}
    \centering
   \includegraphics[width=\columnwidth]{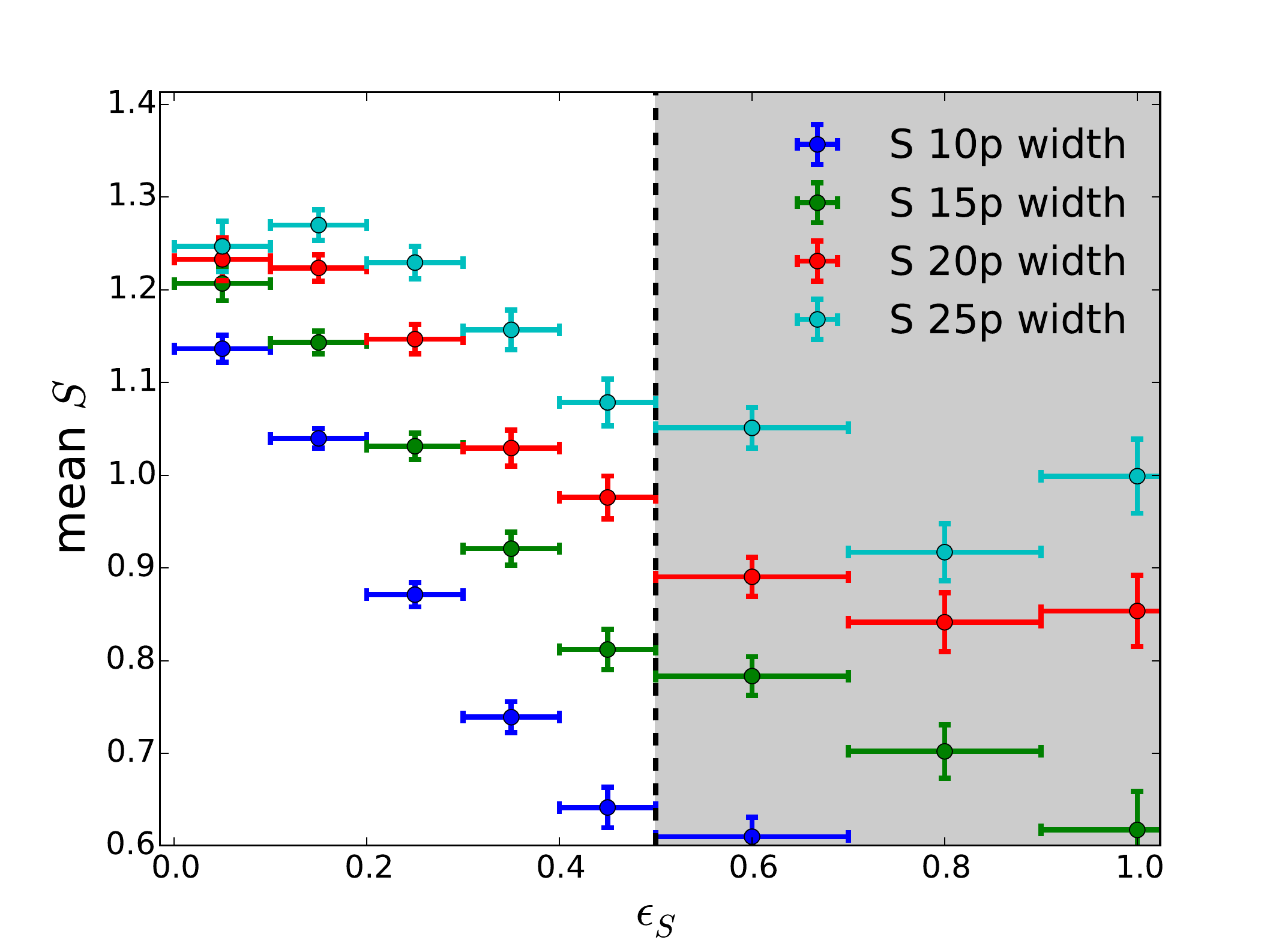}    

	\caption{\label{fig:S-Serr-dependences_wins}
 Mean metal strength as a function of its error, for four different
values of the width of the measurement window: 10, 15, 20 and 25 pixels.
}
 \end{figure}
 
  A second test we do is to examine if the $S$ distribution varies with
the error $\epsilon_S$. In principle, the true
distribution of $S$ should not depend on $\epsilon_S$, which is
determined using only the flux measured in spectral pixels in the
continuum windows and is independent of the flux values in the
measurement window, as explained in \S \ref{subs:continuum}. However,
in reality there is a fairly strong dependence as we show in Figure
\ref{fig:S-Serr_distribution}, where we have divided DLAs into three
intervals of $\epsilon_S$ ([0.0, 0.16), [0.16, 0.33), and [0.33, 0.5]).
If the true $S$ distribution stayed constant, we would expect only an
increased width of the distribution as $\epsilon_S$ increases because
of the convolution with the error, but we also observe a decrease of the
mean value of $S$ with increasing $\epsilon_S$. The effect is nearly the
same for the corrected $S_{\rm HI}$, as seen in the bottom panel.

  This effect is more clearly seen by examining the mean value of $S$ as
a function of $\epsilon_S$, shown in Figure
\ref{fig:S-Serr-dependences}. Our standard upper limit
$\epsilon_S < 0.5$ is shown as the dashed vertical bar, and we divide
the range $0 < \epsilon_S < 0.5$ into 5 intervals. Blue points in the
top panel show that in our standard case, the mean value of $S$
declines from 1.2 to 0.8 as $\epsilon_S$ increases up to its upper
threshold of 0.5. Eliminating DLAs with ${\rm CNR} > 3$ diminishes the
dependence of $S$ on $\epsilon_S$ only slightly, as shown by the red
points in the top panel. This is consistent with the small decrease of
$S$ for ${\rm CNR} < 3$ we saw in Figure \ref{fig:S-CNR}, which cannot
explain the much larger effect of $\epsilon_S$ on the mean $S$. In other
words, the degree of catalogue impurity caused by spectral noise is
relatively small, and the large variation of $S$ with $\epsilon_S$ must
be caused by a different systematic variation of the selected absorption
systems with $\epsilon_S$.

  The green points in the top panel show the visually identified DLAs.
This visual identification \cite{Paris2012} has a larger impact
reducing the dependence of $S$ on $\epsilon_S$. The reason is probably
that the visual identification was influenced by the presence of metal
lines: if metal lines are visually identified, an absorber is more
likely to be flagged as DLA.

  The middle panel of Figure \ref{fig:S-Serr-dependences} shows the
change in the dependence of the mean $S$ with $\epsilon_S$ when we
restrict the sample to DLAs with $N_{\rm HI} > 10^{20.3}\cm^{-2}$
({\it green}), and to $z<3.2$ ({\it red}). As expected, the mean $S$ is
increased for higher column density and lower redshift DLAs. There is
practically no change in the fractional amount by which the mean $S$
drops with $\epsilon_S$ when eliminating the $z<3.2$ DLAs, and a more
appreciable change when the low column density DLAs are eliminated.
However, the strong decline of the mean $S$ with $\epsilon_S$ is
similarly present in all these subsets.
The bottom panel compares the corrected $S_{\rm HI}$
dependence on $\epsilon_{\rm SHI}$ ({\it green}), with the previous
standard uncorrected case. The correction tends to slightly
decrease this dependence.

  As we now show, the principal cause of the decline of the mean $S$
with $\epsilon_S$ is that the DLA redshifts are subject to large errors
when no metal lines are identified. The \lya profile of the DLA is often
contaminated by neighboring \lya forest absorption, which may shift the
best fit of the DLA redshift. Metal lines, which are narrow, usually pin
down the redshift much more accurately. When there is a large redshift
error, the metal lines may be partly shifted outside our measurement
window, reducing the equivalent width that we compute. This is shown in
Figure \ref{fig:S-Serr-dependences_wins}, where the mean $S$ dependence
on $\epsilon_S$ is plotted for four widths of the measurement window:
10, 15, 20 and 25 pixels. The drop in the mean $S$ with $\epsilon_S$ is
gradually reduced with the measurement window width. For a 25 pixel
width, this drop is reduced to only $\sim 10$ \% for $\epsilon_S < 0.5$.
The redshift error required to shift a metal line by half of the 25
pixel width is $\sim 850\kms$, so from the difference in the results for
the 20 and 25 pixels, we conclude that redshift errors can reach this
value for $\sim 10$\% of the DLAs. The remaining dependence can be
explained by a small fraction of even larger redshift errors and the
increase of the mean $S$ with CNR seen in Figure \ref{fig:S-CNR}.


\subsection{Catalogue of metal lines and DLA metal strength}
\label{subs:catalog}

  We have produced a new catalogue of the DLAs in DR12-DLA, including
the metal line equivalent widths computed with the method described in
\S \ref{sec:methodology}. An example of the entries in this catalogue is
presented for three DLAs in Table \ref{tab_catalog}: we list the quasar
spectrum identification and quasar redshift, the CNR, the DLA redshift
and column density, the derived value of $S$ and its error $\epsilon_S$,
the same values corrected for $N_{\rm HI}$ as explained in \S
\ref{subs:corr}, and the equivalent width and errors of each of the 17
metal lines. When lines are not measured for the reasons described in \S
\ref{subs:select_met}, the entries for $W$ and its error are set to
zero. We make the catalogue publicly available  (with metal equivalent widths or not) can be found
on the GitHub repository
\url{https://github.com/andreuandreu/Catalogue_Metal_Strength_DLAs_SDSS-BOSS_DR12}.

\section{Conclusions }
\label{sec:conclude}

 ~\par We have presented a way to classify DLAs according to the strength
of their metal lines. A total of 17 metal lines are used to obtain a
metal strength parameter that is optimized to be measured with the best
possible accuracy. This allows a DLA classification from spectra in
surveys with large numbers of objects, where the spectral
signal-to-noise is often poor. A catalogue of this parameter and several
metal equivalent widths in the DR12-DLA catalogue previously constructed
from the BOSS survey using the method of \cite{Noterdaeme2009} is made
publicly available.

  Our main motivation to present this catalogue is to allow for studies
of the dependence of any property of DLAs as a function of the metal
strength $S$, which after correcting for its dependence on $N_{\rm HI}$
to obtain the new parameter $S_{\rm HI}$, should be a function of the
metal abundance and velocity dispersion of the absorbing gas. We plan
to do two studies along these lines: first, measuring the bias factor
with the technique of \cite{FontRibera2012,Perez2018} applied to subsamples of
the DLAs with different values of $S$ or $S_{\rm HI}$, and measuring the
mean stacked absorption spectrum of the DLA metal lines with the
technique of \cite{lluis2014} for different values of $S$ or $S_{\rm HI}$.

  Our catalogue presents individual equivalent width measurements,
allowing the user to build other combinations of metal line
strengths. One of the applications should be to investigate correlations
among the strength of equivalent widths of different metal lines.
Although the noise is generally very large for individual DLAs, the
large size of the sample can allow for detailed studies.

  An important limitation in using this catalogue is its imperfect
purity. Some of the DLAs may arise due to spectral noise, and some
others may be absorbers of low column density with an absorption
profile that arises from a velocity distribution of absorbing gas in the
\lya forest but is consistent with
a DLA. This can introduce spurious variations of any property we wish
to measure as a function of $S$, if the catalogue purity varies with $S$.
The fact that the mean $S$ as a function of CNR is constant within 5\%
at $CNR>3$ suggests that spectral noise does not introduce substantial
purity variations at $CNR>3$. We measure a dependence of $S$ on
$\epsilon_S$ which we find is mostly due to large redshift errors that
are not corrected when no metal lines are significantly detected, which
cause the metal lines to be partially shifted out of our measurement
window. The impact this may have when measuring the dependence of any
DLA property on the $S$ parameter needs to be borne in mind in future
studies.


\begin{table*}

\centering
\caption{ \label{tab_catalog} Sample of the catalogue for 3 DLAs, with some the information presented:
  \textit{first column} redshift of the Quasar, \textit{second column} Julian(day)-Fiver(number)-Plate(number)
 from the SDSSIII-BOSS numbering. \textit{Third column}  the DLA redshift. \textit{Forth column} density
 as measured by \protect\cite{Noterdaeme2009}. \textit{Fifth column}  is the visual flag of the DLA . 
 In \textit{columns seven to ten} are our corrected  and non corrected \emph{ metal strength} ($S_{\rm HI}$, $S$)
  and their errors  ($\epsilon_{S \rm HI}$ and $\epsilon_S$  as measured in our work. 
\textit{Columns eleven to thirty-three} of this table present the measured $W$ and its standard deviation $\epsilon$ 
for the  sixteen  of the seventeen lines (Mg$\,$I$\,2853$ is eluded here for presentation reasons).
 When 0 is expressed for the $W$ and $\epsilon$  the line has not been selected due to one of the reasons
  described in \S \ref{subs:select_met}.
 Other spectrum characteristics are included in the public catalogue but are not shown here for brevity, 
 these are:  Balnicity index, Right Ascension, and Declination, following the nomenclature from \protect\cite{Noterdaeme2009}. 
  \vspace{2 mm} 
}


\begin{tabular}{*{8}c}
\hline		
	JFP	& Qz	&	CNR & DLAz	&	$\log(N_{HI})$	&Visual flag & $S_{\rm HI} \pm \epsilon_{S\rm HI}$  & $S \pm \epsilon_S$	\\
	\hline
	\hline
	
	56604-7167-0290	& 2.72 & 1.9	&	2.49	&	21.09	&  1 &   0.6142 $\pm$ 0.201 &  0.8894  $\pm$  0.291 	\\
	56265-6151-0936	& 2.48 & 5	     &	2.37	&	20.02	&	0 &	0.8148 $\pm$  0.142 &  0.6088 $\pm$  0.108	\\
	56190-6182-0338	& 2.78 & 7.6   &	2.34	&	21.18	&	1 &	2.0299 $\pm$  0.214  & 3.3966 $\pm$  0.347	\\
\hline		 
 \noalign{\vskip 1mm}

Si$\,$II$\,$1260	&	$\epsilon$(1260) & OI-SiII$\,$1303	&	$\epsilon$(1303)	&	C$\,$II$\,$1334	&	$\epsilon$(1334)	&	Si$\,$II$\,$1526	&	$\epsilon$(1526) 	\\
	\hline
	\hline
				
	 0     &     0  &   0.867 &   0.263 & 0.0989 &   0.424  &  0.8622 &  0.672 \\
	 0      &     0  &   0.150 &    0.18   &  -0.0968  &  0.182  &  0.2600  &  0.15  \\
	 0     &     0   &  0        &     0       &  0      &     0       &  1.467 &   0.151  \\

\hline	
 \noalign{\vskip 1mm} 
								
Fe$\,$II$\,$1608	&	$\epsilon$(1608)	& Al$\,$II$\,$1670	&	$\epsilon$(1670)	&	Si$\,$II$\,$1808	&	$\epsilon$(1808) &  Al$\,$II1854	&	$\epsilon$(1854) 	 	\\
	\hline
	\hline
	
	-1.35  &    1.93 & -0.791   &   0.858 &   0       &      0       & -0.667  &    1.12   \\
	0.260 &   0.150   & 0.078 &  0.202 & -0.320  &  0.27    & 0.240   &  0.27       \\
	 0.362 &   0.172 &  0        &      0    &  0.167    &   0.214  &   0.222 &   0.214  \\

\hline	
 \noalign{\vskip 1mm} 
								
		Al$\,$II$\,$1862	&	$\epsilon$(1862) & Fe$\,$II$\,$2344 	&	$\epsilon$(2344) &  Fe$\,$II$\,$2374	&	$\epsilon$(2374)	&	Fe$\,$II$\,$2382	&	$\epsilon$(2382)		\\
	\hline
	\hline
	  0       &     0    &  -2.94 &    5.69 &   0  &      0  & 0     &    0 \\
	  0.381 &    0.268  &   0    &     0    &   0     &    0  &  0    &   0    \\
	   0      &    0      &   0   &     0    &   0    &    0  & 0     &  0  \\

\hline		
 \noalign{\vskip 1mm} 						
	Fe$\,$II$\,$2587	&	$\epsilon$(2587) & Fe$\,$II$\,$2600	&	$\epsilon$(2600)	&	Mg$\,$II$\,$2796	&	$\epsilon$(2796)	&	Mg$\,$II$\,$2803	&	$\epsilon$(2803)	\\
	\hline
	\hline
  	0        &         0 &     0    &     0  &   0   &     0  &   0   &   0    \\
	0.605 &  0.694 &    0    &     0  &   0    &     0 &   0   &   0  \\ 
  	0        &    0        &   0    &     0  &   0    &     0  &   0   &  0  \\
	\hline	
	
	\end{tabular}

\end{table*}


\section*{Acknowledgements}

 ~\par We would like to thank George Beker for invaluable guidance and help and Andreu
  Font-Ribera and Mat Pieri for useful
discussions . This work has been supported in part by Spanish grants
AYA-2012-33938 and AYA 2015-71091-P.

  Funding for SDSS-III has been provided by the Alfred P. Sloan Foundation, the Participating
Institutions, the National Science Foundation, and the U.S. Department of Energy
Office of Science. The SDSS-III web site is \href{http://www.sdss3.org/}{http://www.sdss3.org/}.

  SDSS-III is managed by the Astrophysical Research Consortium for the Participating
Institutions of the SDSS-III Collaboration including the University of Arizona, the Brazilian
Participation Group, Brookhaven National Laboratory, University of Cambridge, Carnegie
Mellon University, University of Florida, the French Participation Group, the German Participation
Group, Harvard University, the Instituto de Astrofisica de Canarias, the Michigan
State/Notre Dame/JINA Participation Group, Johns Hopkins University, Lawrence Berkeley
National Laboratory, Max Planck Institute for Astrophysics, Max Planck Institute for
Extraterrestrial Physics, New Mexico State University, New York University, Ohio State University,
Pennsylvania State University, University of Portsmouth, Princeton University, the
Spanish Participation Group, University of Tokyo, University of Utah, Vanderbilt University,
University of Virginia, University of Washington, and Yale University.

\newpage

\bibliographystyle{mnras}

\bibliography{ms_DLA}\label{References}



\end{document}